\documentclass[lettersize,journal]{IEEEtran}
\usepackage{cite}

\usepackage{algorithmic}
\usepackage{array}
\usepackage{textcomp}
\usepackage{stfloats}
\usepackage{url}
\usepackage{verbatim}
\usepackage{graphicx}
\usepackage{subfigure}
\usepackage[tbtags]{amsmath}
\usepackage{amsthm,amssymb,bm}
\usepackage{mathrsfs}
\usepackage{enumitem}
\usepackage{tabularx}
\newtheorem{corollary}{Corollary}
\usepackage{threeparttable} % 在导言区引入宏包
\usepackage{nicematrix} % 加载 nicematrix 宏包
\usepackage{tikz}
\usepackage[colorlinks,bookmarksopen,bookmarksnumbered,citecolor=blue, linkcolor=blue, urlcolor=blue]{hyperref}

\usepackage{booktabs}
\usepackage{multirow}
\usepackage{extarrows}
\usepackage[linesnumbered,ruled,vlined,lined,boxed,commentsnumbered]{algorithm2e}
\usepackage{makecell}
\usepackage{xcolor}
\usepackage{balance}
\def\BibTeX{{\rm B\kern-.05em{\sc i\kern-.025em b}\kern-.08em
		T\kern-.1667em\lower.7ex\hbox{E}\kern-.125emX}}

\SetKwComment{Comment}{/* }{ */}
\newtheorem{lemma}{Lemma}
\newtheorem{theorem}{Theorem}

\begin{document}
	% \title{A DSP-oriented Transceiver for ISAC with Random Communication Signals}
	\title{Nonconstant-Modulus Constellations Are Good Enough for Sensing with OFDM Waveforms}
	\author{Junjie~Chen
	\thanks{Manuscript created March 2026; This work was supported by the  (No.xxxxxx). (\textit{Corresponding author: xx~xx.})\\
	Junjie~Chen is with the School of Computing and Artificial Intelligence, Jiangxi University of Finance and Economics, Nanchang 330013, China (e-mail: junjiechen@jxufe.edu.cn).
	}}
	\markboth{IEEE Transactions of xxxx, VOL. xx, NO. xx, NOVEMBER 2026}%
	{How to Use the IEEEtran \LaTeX \ Templates}
	\maketitle

\newtheorem{proposition}{Proposition}
\begin{abstract}
This paper investigates how communication constellations affect joint delay and Doppler estimation under cyclic-prefix orthogonal frequency-division multiplexing (CP-OFDM). Starting from the continuous-time sensing model, we derive an equivalent discrete observation model and specify when it applies. We factorize the Fisher information matrix~(FIM) to separate deterministic channel derivatives from random symbol-energy weights. This structure yields a Jensen-type lower bound on the average conditional Cram\'er--Rao bound~(CRB) for constellations with the same average symbol energy. Although constant-modulus constellations such as phase-shift keying (PSK) attain this lower bound, our two-sided bounds show that the relative CRB gap between quadrature amplitude modulation (QAM) and PSK is of order $Q/(NM)$ for both range and velocity estimation, where $N$, $M$, and $Q$ denote the numbers of subcarriers, OFDM symbols, and targets, respectively. Monte Carlo simulations confirm this scaling and show that the relative gap is already negligible in larger $NM/Q$ regime. This suggests that higher-order QAM can achieve higher communication rates while nearly matching the estimation accuracy of constant-modulus signaling. The mean squared errors~(MSE) of the maximum-likelihood~(ML) estimator approach the corresponding CRBs in simulation, corroborating the derivation. Sliding-subarray multiple signal classification also achieves MSEs close to the corresponding CRBs and substantially outperforms those of on-grid two-dimensional fast Fourier transform~(FFT) estimation, which exhibits grid-induced error floors.
\end{abstract}


\section{Introduction}
\label{sec:introduction}

Integrated sensing and communications (ISAC) is envisioned as an important capability of sixth-generation wireless networks, enabling communication and sensing to share spectrum, hardware, and transmitted waveforms~\cite{liu2022integrated,liu2025sensing}. 
One of the fundamental challenges in ISAC is to design dual-functional waveforms that can simultaneously  perform accurate target sensing and achieve reliable communications. 
Existing waveform design strategies can be classified into three categories: sensing-centric~\cite{hassanien2015dual}, communication-centric~\cite{he2024dual,li2024mimo,li2026ofdm}, and joint design approaches~\cite{chiriyath2017radar,liu2018toward,liu2021cramer}.
Among these, communication-centric ISAC is one of the most promising approaches. On the one hand, it can reuse existing network infrastructure and cellular protocols with only minor modifications, thereby facilitating practical deployment. On the other hand, it exploits data-bearing communication signals for sensing, reducing the need for dedicated sensing transmissions and additional spectrum resources.
In particular, orthogonal frequency-division multiplexing (OFDM) is attractive in this context because its multicarrier structure supports data transmission and provides a convenient representation for sensing~\cite{braun2014ofdm}.

An OFDM-based waveform design using null-space sensing precoding was proposed~\cite{he2024dual} to enhance sensing performance without degrading communication performance. Symbol-level precoding was used to suppress range-Doppler sidelobes in MIMO-OFDM while satisfying multiuser communication quality-of-service requirements~\cite{li2024mimo}. For target echoes with delays exceeding the cyclic prefix (CP) duration, the resulting inter-symbol and inter-carrier interference was characterized, and successive interference cancellation methods were developed to mitigate its impact on range and velocity estimation~\cite{li2026ofdm}.
The CRB-rate tradeoff has also been studied for bistatic ISAC with deterministic sensing signals and Gaussian information symbols, assuming that only the covariance of the information symbols is known to the sensing receiver~\cite{song2026crb}.

In general, radar sensing favors deterministic waveforms with predictable correlation properties, whereas communication relies on random symbols to convey information. These competing requirements give rise to the deterministic-random tradeoff (DRT) between sensing and communications~\cite{xiong2023fundamental}. Symbol randomness affects sensing even when the transmitted realization is known to the sensing receiver. Indeed, precoding designs that explicitly account for random communication data can outperform designs based on a deterministic approximation of the signal sample covariance~\cite{lu2024random}.
For CP-equipped orthogonal waveforms, OFDM has been shown to achieve the lowest average ranging sidelobes under quadrature amplitude modulation (QAM) and phase-shift keying (PSK) constellations~\cite{liu2025cpofdm}. These findings were further extended to pulse-shaped signals~\cite{liu2025uncovering}. With Nyquist pulse shaping, OFDM also achieves the lowest average ranging sidelobe level at every lag for both QAM and PSK. The average squared autocorrelation function (ACF) consists of a pulse-dependent squared mean, termed the ``iceberg,'' and a variance due to data randomness, termed the ``sea level''. Under OFDM, the constant symbol energy of PSK eliminates this variance term, yielding lower ranging sidelobes than nonconstant-modulus QAM. For OFDM, probabilistic constellation shaping provides a means of controlling the sensing-communication tradeoff through the fourth moment of the constellation amplitudes~\cite{du2024reshaping}.
Collectively, these studies establish that modulation, constellation statistics, and pulse shaping affect the sensing properties of random communication waveforms in different ways.

The two-dimensional ambiguity-function analysis provides a particularly useful model for studying delay and Doppler jointly~\cite{zhang2025discrete}. It distinguishes the discrete periodic ambiguity function from the fast-slow-time ambiguity function, the latter relying on a small intra-block Doppler approximation. Under this approximation, delay acts as a circular shift along fast time, whereas Doppler produces phase rotations along slow time. The analysis establishes constellation-dependent sidelobe properties under general orthogonal modulation bases. However, sidelobe optimality does not by itself establish optimality with respect to a joint estimation CRB, which depends on the derivatives of the observation mean and the coupling among all unknown target parameters~\cite{kay1993statistical}.
For multi-target ranging with unknown complex gains, CP-OFDM with PSK symbols has been shown to attain the Jensen lower bound on the average conditional ranging CRB~\cite{liu2026cp}. The same study also shows that, under QAM signaling, CP-OFDM achieves a lower ranging CRB than frequency-spread orthogonal waveforms in the large-sample regime.

In joint delay--Doppler estimation with CP-OFDM, different constellations can produce different symbol-energy fluctuations across a finite frame even at the same average symbol energy. Each symbol's energy weights the Fisher information contributed by the corresponding observation. The numbers of subcarriers and OFDM symbols determine how many observations are available, and the number of targets determines how many parameters are estimated jointly. The resulting CRBs provide a benchmark for evaluating practical algorithms, whose accuracy can also be affected by grid quantization and noise amplification during symbol removal. These considerations motivate questions for a given CP-OFDM waveform. \textit{Do constant-modulus constellations still minimize the average conditional CRB for joint delay-Doppler estimation? How does the gap between the delay-Doppler CRBs of different constellations depend on the numbers of subcarriers, OFDM symbols, and targets? How do different communication constellations perform when used with practical range and velocity estimation algorithms?}

We show that constant-modulus constellations remain CRB-optimal for joint delay-Doppler estimation and quantify the relative CRB gap between QAM and PSK. The gap is already negligible in the typical finite-frame configurations. The simulations also show that ML estimation approaches the corresponding CRBs under both constellations.
The main contributions are as follows.
\begin{enumerate}
	\item We show that sampling the continuous-time CP-OFDM sensing model naturally yields an equivalent discrete-time sensing model. The derivation explicitly identifies the conditions under which range migration and intra-symbol Doppler can be neglected. We also connect three equivalent signal representations through Fourier transforms along fast and slow time, providing insight into how propagation delay and Doppler shift manifest as phase progressions or circular shifts across domains.
	
	\item We derive the FIM for joint estimation of the target delays, Doppler shifts, and complex gains, and reveal a structural factorization that separates the deterministic Jacobian of the sensing response with respect to these parameters from the random symbol-energy weights. This factorization yields a Jensen-type lower bound on the average conditional CRB, which is exactly achieved by constant-modulus constellations such as PSK.
	
	\item We establish finite-frame upper and lower bounds on the relative CRB gap between QAM and PSK for both range and velocity estimation. Under suitable regularity conditions, these bounds show that the gap is of order $Q/(NM)$, where $N$ is the number of subcarriers, $M$ is the number of OFDM symbols, and $Q$ is the number of targets. Monte Carlo simulations corroborate the predicted scaling across different frame dimensions and target counts.
	
	We show that, for a single target under constant-modulus signaling, ML estimation is equivalent to 2D FFT peak search without the grid constraint. Numerical results show that continuous ML closely approaches the corresponding CRBs under both constellations. The on-grid FFT estimator exhibits error floors for off-grid targets. In the scenarios considered, sliding-subarray MUSIC achieves lower estimation errors than FFT, and its estimation errors show a more pronounced dependence on constellation choice than the corresponding CRBs.
\end{enumerate}

The remainder of this paper is organized as follows. Section~\ref{sec:continuous} develops the continuous-time and discrete-time OFDM sensing models. Section~\ref{sec:multicrb} derives the FIM and CRBs, establishes constant-modulus optimality, gives single-target closed forms, and quantifies the relative CRB gap between QAM and PSK. Section~\ref{sec:fft_ml} presents ML estimation for delay and Doppler. Section~\ref{sec:numerical} presents numerical results, and Section~\ref{sec:conclusion} concludes the paper.

\emph{Notation:} Bold uppercase and lowercase letters denote matrices and vectors, respectively. The superscripts $(\cdot)^T$, $(\cdot)^H$, and $(\cdot)^*$ denote transpose, Hermitian transpose, and element-wise conjugation. The operators $\otimes$, $\odot$, and $\operatorname{vec}(\cdot)$ denote the Kronecker product, Hadamard product, and column-wise vectorization. The operators $\operatorname{diag}(\cdot)$, $\operatorname{Tr}(\cdot)$, and $\operatorname{Re}\{\cdot\}$ denote diagonalization, trace, and real part. 

\section{Sensing Signal Model and Insights}
\label{sec:continuous}

\subsection{Continuous-Time Transmit and Receive Signals}

Consider a monostatic ISAC system that uses a CP-OFDM frame with $N$ subcarriers and $M$ OFDM symbols to communicate and sense $Q$ point targets. Let $S[k,m]$ denote the data symbol transmitted on subcarrier $k$ of OFDM symbol $m$, where $0\leq k<N$ and $0\leq m<M$. For each transmission, the symbols are drawn independently from the constellations such as PSK/QAM. The subcarrier spacing is $\Delta f$, the bandwidth is $B=N\Delta f$, and the useful symbol duration is $T_u=1/\Delta f$. With a cyclic prefix (CP) of duration $T_{\mathrm{cp}}$, the total symbol duration is $T_{\mathrm{sym}}=T_u+T_{\mathrm{cp}}$. Taking the start of the first useful symbol as the time origin, the transmitted baseband waveform is
\begin{equation}
	\begin{aligned}
		s_{\mathrm{BB}}(t)
		={}&\frac{1}{\sqrt{N}}\sum_{m=0}^{M-1}\sum_{k=0}^{N-1}S[k,m]e^{j2\pi k\Delta f(t-mT_{\mathrm{sym}})}\\
		&\times\operatorname{rect}\!\left(\frac{t-mT_{\mathrm{sym}}+T_{\mathrm{cp}}}{T_{\mathrm{sym}}}\right),
	\end{aligned}
	\label{eq:frame_waveform}
\end{equation}
where $\operatorname{rect}(x)=1$ for $0\leq x<1$ and is zero otherwise. Each OFDM symbol is prepended with a CP whose duration is no shorter than the maximum round-trip delay.

Let $R_q$, $v_q$, and $\beta_q$ denote the range, radial velocity, and complex reflection coefficient of target $q$, respectively. Its  round-trip delay and Doppler frequency are $\bar{\tau}_q=2R_q/c$ and $\bar{f}_{D,q}=2v_q/\lambda$, respectively, where $c$ is the propagation speed and $\lambda=c/f_c$ is the carrier wavelength. The instantaneous delay satisfies $\bar{\tau}_q(t)\approx\bar{\tau}_q-2v_qt/c$. 
After coherent downconversion, the received baseband signal is
\begin{equation}
	\begin{aligned}
		y_{\mathrm{BB}}(t)
		&=\sum_{q=1}^{Q}\beta_q s_{\mathrm{BB}}\!\left(t-\bar{\tau}_q(t)\right)e^{-j2\pi f_c\bar{\tau}_q(t)}+w(t)\\
		&\overset{(a)}{\approx}\sum_{q=1}^{Q}\beta_q s_{\mathrm{BB}}(t-\bar{\tau}_q)e^{-j2\pi f_c\bar{\tau}_q(t)}+w(t)\\
		&\overset{(b)}{\approx}\sum_{q=1}^{Q}\beta_q e^{-j2\pi f_c\bar{\tau}_q}s_{\mathrm{BB}}(t-\bar{\tau}_q)e^{j2\pi\bar{f}_{D,q}t}+w(t)\\
		&\overset{(c)}{=}\sum_{q=1}^{Q}\alpha_q s_{\mathrm{BB}}(t-\bar{\tau}_q)e^{j2\pi\bar{f}_{D,q}t}+w(t),
	\end{aligned}
	\label{eq:continuous_delay_doppler}
\end{equation}
where $w(t)$ is the complex baseband noise. Step~$(a)$ neglects range migration under $B|\bar{\tau}_q(t)-\bar{\tau}_q|\ll1$ throughout the OFDM frame~\cite{bok2021movement,du2026filtering}. Thus, the time-varying delay in the signal envelope is replaced by its reference value~\cite{hakobyan2016narrowband}. Step~$(b)$ follows from~$e^{-j2\pi f_c\bar{\tau}_q(t)} \approx e^{-j2\pi f_c\bar{\tau}_q}e^{j2\pi\bar{f}_{D,q}t}$.
Step~$(c)$ absorbs the constant carrier phase into the effective complex gain $\alpha_q\triangleq\beta_q e^{-j2\pi f_c\bar{\tau}_q}$.

Assuming perfect timing synchronization between the receiver and the transmitted frame, the $m$th useful receive window starts at $mT_{\mathrm{sym}}$. Substituting~\eqref{eq:frame_waveform} into~\eqref{eq:continuous_delay_doppler} and observing the received signal within this window gives, for $0\leq u<T_u$,
\begin{equation}
	\begin{aligned}
		&r_m(u) \triangleq y_{\mathrm{BB}}(mT_{\mathrm{sym}}+u)\\
		&\quad\approx \sum_{q=1}^{Q}\alpha_q s_{\mathrm{BB}}(mT_{\mathrm{sym}}+u-\bar{\tau}_q)  e^{j2\pi\bar f_{D,q}(mT_{\mathrm{sym}}+u)}+w_m(u)\\
		&\quad\overset{(a)}{=} \frac{1}{\sqrt N}\sum_{q=1}^{Q}\sum_{k=0}^{N-1}\alpha_q S[k,m]  \underbrace{e^{j2\pi u\bar f_{D,q}}}_{\text{\textcircled{\scriptsize 1}\ Intra-symbol Doppler}}\\
		&\qquad \times \underbrace{e^{j2\pi mT_{\mathrm{sym}}\bar f_{D,q}}}_{\text{\textcircled{\scriptsize 2}\ Inter-symbol Doppler}} \underbrace{e^{-j2\pi k\Delta f\bar{\tau}_q}}_{\text{\textcircled{\scriptsize 3}\ Subcarrier delay phase}} \underbrace{e^{j2\pi k\Delta f u}}_{\text{\textcircled{\scriptsize 4}\ Subcarrier basis}} \\
		&\qquad +w_m(u)\\
		&\quad\overset{(b)}{\approx} \frac{1}{\sqrt N}\sum_{q=1}^{Q}\sum_{k=0}^{N-1}\alpha_q S[k,m]e^{j2\pi mT_{\mathrm{sym}}\bar f_{D,q}}e^{-j2\pi k\Delta f\bar{\tau}_q}\\
		&\qquad \times e^{j2\pi k\Delta f u}+w_m(u),
	\end{aligned}
	\label{eq:receive_window_full}
\end{equation}
where $w_m(u)\triangleq w(mT_{\mathrm{sym}}+u)$. In $(a)$, CP coverage ensures that $-T_{\mathrm{cp}}\leq u-\bar{\tau}_q<T_u$, so the delayed waveform belongs to the CP-extended $m$th symbol. In $(b)$, we assume that the fast-time Doppler is sufficiently small, i.e., $2\pi|\bar f_{D,q}|T_u\ll 1$ for all $q$~\cite{zhang2025discrete}. Then $e^{j2\pi u\bar f_{D,q}}\approx 1$ for $0\leq u<T_u$.
Factors \textcircled{\scriptsize 1} and \textcircled{\scriptsize 2} describe the Doppler phase variation within an OFDM symbol and across OFDM symbols, respectively. Factor \textcircled{\scriptsize 3} gives the delay-induced phase shift on the $k$th subcarrier, and factor \textcircled{\scriptsize 4} is the $k$th subcarrier basis function within the useful receive window.

\noindent\textbf{Remark 1 (Constellation moments).}
We consider zero-mean constellations and normalize the average symbol energy to unity, so that $\mathbb{E}[S[k,m]]=0$ and $E_s\triangleq\mathbb{E}[|S[k,m]|^2]=1$. The constellation kurtosis is the normalized fourth central moment
\begin{equation}
	\kappa\triangleq
	\frac{\mathbb{E}[|S[k,m]-\mathbb{E}[S[k,m]]|^4]}
	{(\mathbb{E}[|S[k,m]-\mathbb{E}[S[k,m]]|^2])^2}
	=\mathbb{E}[|S[k,m]|^4].
	\label{eq:constellation_kurtosis}
\end{equation}

\subsection{Discrete-Time Observation Model by Sampling}
The discrete-time observation model can be obtain by sampling the useful receive window in~\eqref{eq:receive_window_full} at $u=n\Delta t$, where $n=0,\ldots,N-1$ and $\Delta t=T_u/N=1/(N\Delta f)$. The CP contains $N_{\mathrm{cp}}$ samples, so $T_{\mathrm{cp}}=N_{\mathrm{cp}}\Delta t$ and $T_{\mathrm{sym}}=(N+N_{\mathrm{cp}})\Delta t$. With delay and Doppler resolutions $\Delta\tau=\Delta t$ and $\Delta f_D=1/(MT_{\mathrm{sym}})$, define the normalized target coordinates as
\begin{equation}
	\tau_q=\frac{\bar{\tau}_q}{\Delta\tau},
	\quad \nu_q=\frac{\bar f_{D,q}}{\Delta f_D}.
	\label{eq:normalized_parameters}
\end{equation}
At $u=n\Delta t$, the four factors in~\eqref{eq:receive_window_full} become \textcircled{\scriptsize 1} $\left.e^{j2\pi u\bar f_{D,q}}\right|_{u=n\Delta t}=e^{j2\pi n\nu_q/[M(N+N_{\mathrm{cp}})]}$, \textcircled{\scriptsize 2} $e^{j2\pi mT_{\mathrm{sym}}\bar f_{D,q}}=e^{j2\pi m\nu_q/M}$, \textcircled{\scriptsize 3} $e^{-j2\pi k\Delta f\bar{\tau}_q}=e^{-j2\pi k\tau_q/N}$, and \textcircled{\scriptsize 4} $\left.e^{j2\pi k\Delta f u}\right|_{u=n\Delta t}=e^{j2\pi kn/N}$.
The discrete-time counterpart of the small intra-symbol Doppler condition in~\eqref{eq:receive_window_full} is
\begin{equation}
	\frac{2\pi N|\nu_q|}{M(N+N_{\mathrm{cp}})}=2\pi|\bar f_{D,q}|T_u\ll 1,\quad q=1,\ldots,Q.
	\label{eq:small_fast_time_doppler}
\end{equation}

We use $\tau_q$ and $\nu_q$ throughout the subsequent analysis. They are integers for on-grid targets and may take noninteger values for off-grid targets. The corresponding range and radial velocity are $R_q=\Delta R\,\tau_q$ and $v_q=\Delta v\,\nu_q$, where $\Delta R=c\Delta\tau/2$ and $\Delta v=\lambda\Delta f_D/2$ are the range and velocity resolutions, respectively.

Under~\eqref{eq:small_fast_time_doppler}, factor \textcircled{\scriptsize 1} is approximated by one over the $N$ useful samples. Sampling the last line of~\eqref{eq:receive_window_full} and combining factors \textcircled{\scriptsize 3} and \textcircled{\scriptsize 4} yields
\begin{equation}
	\begin{aligned}
		Y_r[n,m] &\triangleq r_m(n\Delta t)\\
		&\approx \frac{1}{\sqrt N}\sum_{q=1}^{Q}\sum_{k=0}^{N-1}\alpha_q e^{j2\pi m\nu_q/M}S[k,m]e^{j2\pi k(n-\tau_q)/N}\\
		&\quad +W_r[n,m],
	\end{aligned}
	\label{eq:sampled_receive_model}
\end{equation}
where $W_r[n,m]=w_m(n\Delta t)$.
Let $\boldsymbol F_N$ denote the unitary $N$-point DFT matrix, with $[\boldsymbol F_N]_{k,n}=N^{-1/2}e^{-j2\pi kn/N}$. Sampling the useful part of the transmitted waveform in~\eqref{eq:frame_waveform} gives the fast-time and slow-time matrix
\begin{equation}
	\boldsymbol X=\boldsymbol F_N^H\boldsymbol S\in\mathbb C^{N\times M}.
	\label{eq:ofdm_matrix_modulation}
\end{equation}
Define $\boldsymbol B_N(\tau_q)=\operatorname{diag}_{k=0}^{N-1}(e^{-j2\pi k\tau_q/N})$ and $\boldsymbol D_M(\nu_q)=\operatorname{diag}_{m=0}^{M-1}(e^{j2\pi m\nu_q/M})$. Collecting the samples in~\eqref{eq:sampled_receive_model} gives
\begin{equation}
	\boldsymbol Y_r\approx\sum_{q=1}^{Q}\alpha_q\boldsymbol F_N^H\boldsymbol B_N(\tau_q)\boldsymbol S\boldsymbol D_M(\nu_q)+\boldsymbol W_r,
	\label{eq:sampled_receive_matrix}
\end{equation}
where $\boldsymbol Y_r$ and $\boldsymbol W_r$ collect the useful receive samples and the corresponding noise samples, respectively.

Applying an $N$-point FFT along fast time at the receiver, we obtain the subcarrier-slow-time observation model
\begin{equation}
	\boldsymbol Y_s=\boldsymbol F_N\boldsymbol Y_r\approx\boldsymbol H\odot\boldsymbol S+\boldsymbol W_s,
	\label{eq:frequency_model}
\end{equation}
where $\boldsymbol W_s=\boldsymbol F_N\boldsymbol W_r$ and $\odot$ denotes elementwise multiplication. $\boldsymbol H$~is the sensing-channel matrix
\begin{equation}
	\boldsymbol H=\sum_{q=1}^{Q}\alpha_q\boldsymbol b(\tau_q)\boldsymbol d^H(\nu_q),
	\label{eq:channel_outer}
\end{equation}
where $\boldsymbol b(\tau)=[1,e^{-j2\pi\tau/N},\ldots,e^{-j2\pi(N-1)\tau/N}]^T\in\mathbb C^N$ and $\boldsymbol d(\nu)=[1,e^{-j2\pi\nu/M},\ldots,e^{-j2\pi(M-1)\nu/M}]^T\in\mathbb C^M$ are the delay and Doppler steering vectors, respectively.
The subcarrier-slow-time representation in~\eqref{eq:frequency_model} provides the observation model for the subsequent CRB analysis.
The transmitted symbols $\boldsymbol S$ are known to the monostatic sensing receiver, and the noise is modeled as $\operatorname{vec}(\boldsymbol W_s)\sim\mathcal{CN}(\boldsymbol 0,\sigma_W^2\boldsymbol I_{NM})$.

\subsection{Insights from the Equivalent Signal Representations}

The same sensing process admits several equivalent domain-specific representations, linked by Fourier transforms along fast time and slow time. Each representation exposes a different structure of the target response, providing insight into how delay and Doppler can be estimated. The signal propagating through the physical channel is always a time-domain waveform. As the signal is transformed between domains at different stages of OFDM processing, the same effects of propagation delay and Doppler shift take different mathematical forms, typically appearing as phase rotations or circular shifts.

\subsubsection{Fast-Time and Slow-Time Representation}

For an integer delay $\ell$, the corresponding circular-shift matrix is
\begin{equation}
	\boldsymbol J_N(\ell)=
	\begin{bmatrix}
		\boldsymbol 0_{\ell\times(N-\ell)} & \boldsymbol I_\ell\\
		\boldsymbol I_{N-\ell} & \boldsymbol 0_{(N-\ell)\times\ell}
	\end{bmatrix},
	\quad 1\leq\ell<N,
	\label{eq:integer_circular_delay}
\end{equation}
with $\boldsymbol J_N(0)=\boldsymbol I_N$. Since $\boldsymbol J_N(\ell)$ is circulant, it admits the diagonalization
\begin{equation}
	\boldsymbol J_N(\ell)=\boldsymbol F_N^H\boldsymbol B_N(\ell)\boldsymbol F_N.
	\label{eq:circular_delay_diagonalization}
\end{equation}

The same Fourier representation defines $\boldsymbol J_N(\tau)$ for fractional delays. Using~\eqref{eq:circular_delay_diagonalization} and $\boldsymbol X=\boldsymbol F_N^H\boldsymbol S$, we write~\eqref{eq:sampled_receive_matrix} as
\begin{equation}
	\boldsymbol Y_r\approx\sum_{q=1}^{Q}\alpha_q\boldsymbol J_N(\tau_q)\boldsymbol X\boldsymbol D_M(\nu_q)+\boldsymbol W_r.
	\label{eq:time_domain_circular_model}
\end{equation}
For target $q$, left multiplication by $\boldsymbol J_N(\tau_q)$ circularly delays the fast-time waveform in each column of $\boldsymbol X$. Right multiplication by $\boldsymbol D_M(\nu_q)$ multiplies column $m$ by $e^{j2\pi m\nu_q/M}$, producing a phase progression across OFDM symbols. Correlating $\boldsymbol X$ with its circularly delayed and Doppler-shifted replicas gives the time-domain ambiguity function. Using the received waveform $\boldsymbol Y_r$ in this correlation gives a matched-filter response over candidate delays and Doppler shifts.
\textit{The locations of the peaks in this matched-filter response provide the delay and Doppler estimates.}

Applying $\operatorname{vec}(\boldsymbol A\boldsymbol X\boldsymbol B)=(\boldsymbol B^T\otimes\boldsymbol A)\operatorname{vec}(\boldsymbol X)$ to~\eqref{eq:time_domain_circular_model} and using $\boldsymbol D_M^T(\nu_q)=\boldsymbol D_M(\nu_q)$ yields
\begin{equation}
	\boldsymbol y_r\approx\sum_{q=1}^{Q}\alpha_q\bigl(\boldsymbol D_M(\nu_q)\otimes\boldsymbol J_N(\tau_q)\bigr)\boldsymbol x+\boldsymbol w_r,
	\label{eq:vectorized_fast_slow_observation}
\end{equation}
where $\boldsymbol y_r=\operatorname{vec}(\boldsymbol Y_r)$, $\boldsymbol x=\operatorname{vec}(\boldsymbol X)$, and $\boldsymbol w_r=\operatorname{vec}(\boldsymbol W_r)$. This is exactly the observation model in~\cite[eq.~(37)]{zhang2025discrete}. Equation~\eqref{eq:vectorized_fast_slow_observation} follows from sampling the continuous-time echo within the useful receive windows. The discrete-time derivation in~\cite{zhang2025discrete} composes CP insertion, frame-level delay-Doppler propagation, and CP removal. In both derivations, CP coverage gives a circular delay within each OFDM symbol, and the small intra-symbol Doppler condition gives one Doppler phase per symbol. These two properties yield the same Kronecker product of the fast-time delay and slow-time Doppler operators.

\subsubsection{Subcarrier and Slow-Time Representation}
\begin{figure}[t]
	\centering
	\includegraphics[width=1\linewidth]{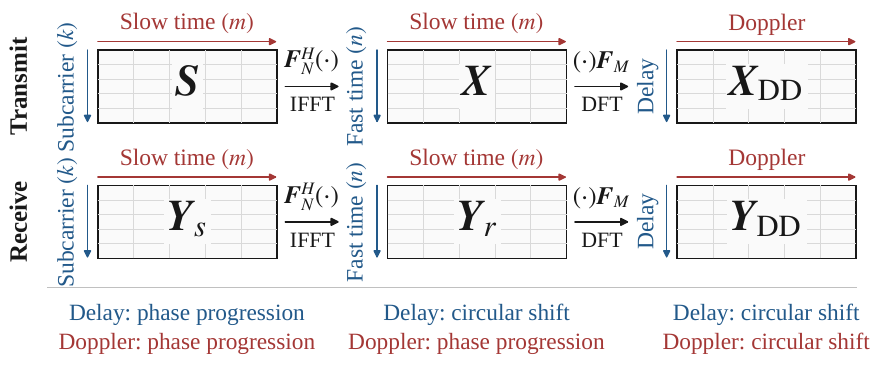}
	\caption{Domain transformations and representations of delay-Doppler effects at different processing stages.}
	\label{fig:signal_domains}
\end{figure}
Left-multiplying~\eqref{eq:sampled_receive_matrix} by $\boldsymbol F_N$ gives
\begin{equation}
	\boldsymbol Y_s=\boldsymbol F_N\boldsymbol Y_r\approx\sum_{q=1}^{Q}\alpha_q\boldsymbol B_N(\tau_q)\boldsymbol S\boldsymbol D_M(\nu_q)+\boldsymbol W_s.
	\label{eq:subcarrier_slow_time_model}
\end{equation}
The fast-time circular delay becomes a phase progression across subcarriers, while the slow-time Doppler phases are preserved. The two phase progressions form the delay and Doppler steering vectors in~\eqref{eq:channel_outer}, which provide a separable parameterization of each target response for joint estimation.
\textit{These steering vectors can be used for joint maximum-likelihood estimation directly from $bm{Y}_s$ or for 2D MUSIC after symbolwise division.}

\subsubsection{Delay-Doppler Representation}

Slow time and Doppler form another Fourier-dual pair. Let $\boldsymbol F_M$ denote the unitary $M$-point DFT matrix. Applying the DFT along slow time gives
\begin{equation}
	\begin{aligned}
		\boldsymbol X_{\mathrm{DD}} &= \boldsymbol X\boldsymbol F_M=\boldsymbol F_N^H\boldsymbol S\boldsymbol F_M,\\
		\boldsymbol Y_{\mathrm{DD}} &= \boldsymbol Y_r\boldsymbol F_M=\boldsymbol F_N^H\boldsymbol Y_s\boldsymbol F_M.
	\end{aligned}
	\label{eq:waveform_dd_representations}
\end{equation}
Since $\boldsymbol J_M(-\nu)=\boldsymbol F_M^H\boldsymbol D_M(\nu)\boldsymbol F_M$, right multiplication of~\eqref{eq:time_domain_circular_model} by $\boldsymbol F_M$ yields
\begin{equation}
	\boldsymbol Y_{\mathrm{DD}}\approx\sum_{q=1}^{Q}\alpha_q\boldsymbol J_N(\tau_q)\boldsymbol X_{\mathrm{DD}}\boldsymbol J_M(-\nu_q)+\boldsymbol W_{\mathrm{DD}},
	\label{eq:dd_waveform_model}
\end{equation}
where $\boldsymbol W_{\mathrm{DD}}=\boldsymbol W_r\boldsymbol F_M$. Left multiplication by $\boldsymbol J_N(\tau_q)$ shifts each column along the delay axis, and right multiplication by $\boldsymbol J_M(-\nu_q)$ shifts each row along the Doppler axis. Each target echo therefore has the form of a two-dimensional circularly shifted transmit waveform.
\textit{The delays and Doppler shifts can be estimated jointly by fitting a superposition of fractionally shifted replicas of $\boldsymbol{X}_{\mathrm{DD}}$ to $\boldsymbol{Y}_{\mathrm{DD}}$ in the least-squares sense.}
Fig.~\ref{fig:signal_domains} shows how Fourier transforms relate the three waveform representations and how delay and Doppler appear in each.

\subsection{Two-Dimensional Delay-Doppler Channel Impulse Response}

Applying an inverse DFT along the subcarrier dimension of $\boldsymbol H$ and a DFT along its slow-time dimension yields the sparsely two-dimensional channel impulse response (CIR), $\boldsymbol G=\boldsymbol F_N^H\boldsymbol H\boldsymbol F_M$. For on-grid targets, its entries are
\begin{equation}
	G[\ell,p]=\sqrt{NM}\sum_{q=1}^{Q}\alpha_q\,\delta_N[\ell-\tau_q]\,\delta_M[p-\nu_q],~ \tau_q,\nu_q\in\mathbb Z,
	\label{eq:dd_channel_impulse_response}
\end{equation}
where $\ell=0,\ldots,N-1$ and $p=0,\ldots,M-1$ index the delay and Doppler bins, respectively. Each on-grid target contributes a single tap at its delay-Doppler bin.
The reduced observation model in~\eqref{eq:frequency_model} can thus be written as
\begin{equation}
	\boldsymbol Y_s=\bigl(\boldsymbol F_N\boldsymbol G\boldsymbol F_M^H\bigr)\odot\boldsymbol S+\boldsymbol W_s.
	\label{eq:dd_cir_observation}
\end{equation}
With known $\boldsymbol S$, on-grid sensing amounts to recovering the sparse support of $\boldsymbol G$ from $\boldsymbol Y_s$. In noisy observations, candidate targets are identified at bins where the estimated tap power $|\widehat G[\ell,p]|^2$ exceeds a detection threshold.
\textit{For on-grid targets, the delay-Doppler CIR $\boldsymbol{G}$ can be estimated from $\boldsymbol{Y}_s$ by fitting~\eqref{eq:dd_cir_observation} via sparsity-constrained least squares, with the locations of the recovered taps giving the delay and Doppler bins~\cite{oikonomou2026elliptic,knill2018high}.}

\section{The Structure of FIM and CRBs}\label{sec:multicrb}

This section analyzes the joint delay--Doppler estimation CRB in the presence of unknown complex target gains. We first derive a structural factorization of the complete FIM, which provides the basis for the subsequent average conditional CRB analysis.

\subsection{Structural Factorization of the FIM}

We begin with the subcarrier observation model, with the transmitted symbols and the number of targets assumed known. Vectorizing the received matrix in \eqref{eq:frequency_model} yields
\begin{equation}
	\boldsymbol{y}
	=
	\boldsymbol{B}(\boldsymbol{\theta})\boldsymbol{\alpha}
	+\boldsymbol{w},
	\label{eq:linear_observation}
\end{equation}
where $\boldsymbol{y}=\operatorname{vec}(\boldsymbol{Y}_s)$, $\boldsymbol{\alpha}=[\alpha_1,\ldots,\alpha_Q]^T$, and $\boldsymbol{w}=\operatorname{vec}(\boldsymbol{W}_s)\sim\mathcal{CN}(\boldsymbol{0},\sigma_W^2\boldsymbol{I}_{NM})$. The vector $\boldsymbol{\theta}=[\boldsymbol{\tau}^T,\boldsymbol{\nu}^T]^T\in\mathbb{R}^{2Q}$ collects the normalized target delays and Doppler shifts, where $\boldsymbol{\tau}=[\tau_1,\ldots,\tau_Q]^T$ and $\boldsymbol{\nu}=[\nu_1,\ldots,\nu_Q]^T$. The data-weighted steering matrix is
\begin{equation}
	\begin{aligned}
		\boldsymbol{B}(\boldsymbol{\theta})
		&=\operatorname{diag}\!\left(\operatorname{vec}(\boldsymbol{S})\right)
		\boldsymbol{A}(\boldsymbol{\theta})\\
		&=\left[
		\operatorname{vec}(\boldsymbol{S})\odot\boldsymbol{a}_1,
		\ldots,
		\operatorname{vec}(\boldsymbol{S})\odot\boldsymbol{a}_Q
		\right]\in\mathbb{C}^{NM\times Q},
	\end{aligned}
	\label{eq:data_weighted_steering}
\end{equation}
with
\begin{equation}
	\boldsymbol{A}(\boldsymbol{\theta})
	=
	[\boldsymbol{a}_1,\ldots,\boldsymbol{a}_Q]
	\in\mathbb{C}^{NM\times Q},
	\label{eq:full_steering}
\end{equation}
and $\boldsymbol{a}_q=\boldsymbol{d}^*(\nu_q)\otimes\boldsymbol{b}(\tau_q)\in\mathbb{C}^{NM}$ denoting the joint delay-Doppler steering vector of target $q$. Using $\operatorname{vec}(\boldsymbol{b}\boldsymbol{d}^H)=\boldsymbol{d}^*\otimes\boldsymbol{b}$, the vectorized sensing channel satisfies
\begin{equation}
	\boldsymbol{h}
	\triangleq
	\operatorname{vec}(\boldsymbol{H})
	=
	\boldsymbol{A}(\boldsymbol{\theta})\boldsymbol{\alpha}.
	\label{eq:vectorized_sensing_channel}
\end{equation}
The normalized target delays, Doppler shifts, and complex gains are jointly estimated through the real parameter vector
\begin{equation}
	\boldsymbol{\vartheta}
	=
	\left[
	\boldsymbol{\tau}^T,
	\boldsymbol{\nu}^T,
	\boldsymbol{\alpha}_R^T,
	\boldsymbol{\alpha}_I^T
	\right]^T
	\in\mathbb{R}^{4Q},
	\label{eq:parameter_vector}
\end{equation}
where
\begin{equation}
	\begin{aligned}
		\boldsymbol{\alpha}_R
		&=[\operatorname{Re}\{\alpha_1\},\ldots,
		\operatorname{Re}\{\alpha_Q\}]^T,\\
		\boldsymbol{\alpha}_I
		&=[\operatorname{Im}\{\alpha_1\},\ldots,
		\operatorname{Im}\{\alpha_Q\}]^T.
	\end{aligned}
	\label{eq:gain_parameter_vectors}
\end{equation}
We further denote $\boldsymbol{\xi}=[\boldsymbol{\alpha}_R^T,\boldsymbol{\alpha}_I^T]^T$, so that $\boldsymbol{\vartheta}=[\boldsymbol{\theta}^T,\boldsymbol{\xi}^T]^T$. The effective gains are treated as independent unknown complex parameters. Accordingly, the carrier-delay phase absorbed into these gains is not treated as separately available delay information.
Conditioned on the transmitted symbols, the observation is complex Gaussian vector with mean
\begin{equation}
	\boldsymbol{\mu}(\boldsymbol{\vartheta};\boldsymbol{S})
	=
	\boldsymbol{B}(\boldsymbol{\theta})\boldsymbol{\alpha}.
	\label{eq:conditional_mean}
\end{equation}
For notational convenience, we use $\boldsymbol{A}$ and $\boldsymbol{A}(\boldsymbol{\theta})$, as well as $\boldsymbol{B}$ and $\boldsymbol{B}(\boldsymbol{\theta})$, interchangeably hereafter.

Since the noise covariance is independent of $\boldsymbol{\vartheta}$, the $(i,j)$th element of the FIM $\boldsymbol{J}(\boldsymbol{S})\in\mathbb{R}^{4Q\times4Q}$ is
\begin{equation}
	\begin{aligned}
		[\boldsymbol{J}(\boldsymbol{S})]_{i,j}
		&=
		\frac{2}{\sigma_W^2}
		\operatorname{Re}
		\left\{
		\left(
		\frac{\partial\boldsymbol{\mu}}{\partial\vartheta_i}
		\right)^H
		\frac{\partial\boldsymbol{\mu}}{\partial\vartheta_j}
		\right\}\\
		&=
		\frac{2}{\sigma_W^2}
		\operatorname{Re}
		\left\{
		\sum_{\ell=1}^{NM}
		\left(
		\frac{\partial\mu_\ell}{\partial\vartheta_i}
		\right)^*
		\frac{\partial\mu_\ell}{\partial\vartheta_j}
		\right\},
		\quad i,j=1,\ldots,4Q,
	\end{aligned}
\end{equation}
where $\mu_\ell$ denotes the $\ell$th entry of $\boldsymbol{\mu}$. For target $q$, differentiating the mean with respect to its normalized delay and Doppler gives
\begin{equation}
	\begin{aligned}
		\left[
		\frac{\partial\boldsymbol{\mu}}{\partial\tau_q}
		\right]_{k+mN+1}
		&=
		-\frac{j2\pi k}{N}\,
		\alpha_q S[k,m]\,
		e^{-j2\pi k\tau_q/N}
		e^{j2\pi m\nu_q/M},\\
		\left[
		\frac{\partial\boldsymbol{\mu}}{\partial\nu_q}
		\right]_{k+mN+1}
		&=
		\frac{j2\pi m}{M}\,
		\alpha_q S[k,m]\,
		e^{-j2\pi k\tau_q/N}
		e^{j2\pi m\nu_q/M},
	\end{aligned}
\end{equation}
where $k=0,\ldots,N-1$, $m=0,\ldots,M-1$, and the index $k+mN+1$ follows from column-wise vectorization.

To obtain a compact FIM, let $\boldsymbol{K},\boldsymbol{L}\in\mathbb{R}^{NM\times NM}$ collect the subcarrier and OFDM-symbol indices, respectively,
\begin{equation}
	\begin{aligned}
		\boldsymbol{K}
		&=
		\boldsymbol{I}_M\otimes
		\operatorname{diag}(0,\ldots,N-1),\\
		\boldsymbol{L}
		&=
		\operatorname{diag}(0,\ldots,M-1)
		\otimes\boldsymbol{I}_N.
	\end{aligned}
\end{equation}
The delay and Doppler Jacobians of $\boldsymbol{\mu}(\boldsymbol{\vartheta};\boldsymbol{S})$ are
\begin{equation}
	\begin{aligned}
		\boldsymbol{D}_\tau
		&\triangleq
		\frac{\partial\boldsymbol{\mu}}
		{\partial\boldsymbol{\tau}^{T}}
		=
		-\frac{j2\pi}{N}\,
		\boldsymbol{K}\boldsymbol{B}
		\operatorname{diag}(\boldsymbol{\alpha}),\\
		\boldsymbol{D}_\nu
		&\triangleq
		\frac{\partial\boldsymbol{\mu}}
		{\partial\boldsymbol{\nu}^{T}}
		=
		\frac{j2\pi}{M}\,
		\boldsymbol{L}\boldsymbol{B}
		\operatorname{diag}(\boldsymbol{\alpha}).
	\end{aligned}
\end{equation}
Denote $\boldsymbol{D}=[\boldsymbol{D}_\tau,\boldsymbol{D}_\nu]\in\mathbb{C}^{NM\times2Q}$. Since $\boldsymbol{\alpha}=\boldsymbol{\alpha}_R+j\boldsymbol{\alpha}_I$, the Jacobians with respect to the real and imaginary gain components are
\begin{equation}
	\begin{aligned}
		\frac{\partial\boldsymbol{\mu}}
		{\partial\boldsymbol{\alpha}_R^{T}}
		&=
		\frac{\partial
			\left[\boldsymbol{B}(\boldsymbol{\alpha}_R+j\boldsymbol{\alpha}_I)\right]}
		{\partial\boldsymbol{\alpha}_R^{T}}
		=
		\boldsymbol{B},\\
		\frac{\partial\boldsymbol{\mu}}
		{\partial\boldsymbol{\alpha}_I^{T}}
		&=
		\frac{\partial
			\left[\boldsymbol{B}(\boldsymbol{\alpha}_R+j\boldsymbol{\alpha}_I)\right]}
		{\partial\boldsymbol{\alpha}_I^{T}}
		=
		j\boldsymbol{B}.
	\end{aligned}
\end{equation}
Collecting the derivatives gives the complete mean Jacobian
\begin{equation}
	\begin{aligned}
		\boldsymbol{G}_{\mu}
		\triangleq
		\frac{\partial\boldsymbol{\mu}}
		{\partial\boldsymbol{\vartheta}^{T}}
		=
		[\boldsymbol{D}_\tau,\boldsymbol{D}_\nu,\boldsymbol{B},j\boldsymbol{B}]
		\in\mathbb{C}^{NM\times4Q}.
	\end{aligned}
\end{equation}
The complete FIM consequently admits the compact expression
\begin{equation}
	\begin{aligned}
		\boldsymbol{J}(\boldsymbol{S})
		=
		\frac{2}{\sigma_W^2}
		\operatorname{Re}
		\left\{
		\boldsymbol{G}_{\mu}^H\boldsymbol{G}_{\mu}
		\right\}.
	\end{aligned}
	\label{eq:complete_fim}
\end{equation}

To reveal how the target parameters and transmitted symbols contribute to this FIM, we factor the symbol weights out of the mean Jacobian. The diagonal matrices $\boldsymbol{K}$ and $\boldsymbol{L}$ commute with $\operatorname{diag}(\operatorname{vec}(\boldsymbol{S}))$, giving
\begin{equation}
	\boldsymbol{G}_{\mu}
	=
	\operatorname{diag}(\operatorname{vec}(\boldsymbol{S}))
	\boldsymbol{G}_0,
	\label{eq:jacobian_factorization}
\end{equation}
where $\boldsymbol{G}_0\in\mathbb{C}^{NM\times4Q}$ is the Jacobian of the vectorized sensing channel with respect to the complete target-parameter vector, given by
\begin{equation}
	\begin{aligned}
		\boldsymbol{G}_0
		&\triangleq
		\frac{\partial\boldsymbol{h}}
		{\partial\boldsymbol{\vartheta}^T}\\
		&= \left[ -\frac{j2\pi}{N} \boldsymbol{K}\boldsymbol{A} \operatorname{diag}(\boldsymbol{\alpha}),
		\frac{j2\pi}{M} \boldsymbol{L}\boldsymbol{A} \operatorname{diag}(\boldsymbol{\alpha}), \boldsymbol{A}, j\boldsymbol{A} \right].
	\end{aligned}
	\label{eq:channel_jacobian_meaning}
\end{equation}
For fixed target parameters, $\boldsymbol{G}_0$ is deterministic and describes the sensitivity of the sensing channel to changes in the target delays, Doppler shifts, and complex gains. Substituting this factorization into~\eqref{eq:complete_fim} gives the equivalent expression
\begin{equation}
	\boldsymbol{J}(\boldsymbol{S})
	=
	\frac{2}{\sigma_W^2}
	\operatorname{Re}
	\left\{
	\boldsymbol{G}_0^H
	\operatorname{diag}
	\left(
	|\operatorname{vec}(\boldsymbol{S})|^2
	\right)
	\boldsymbol{G}_0
	\right\}.
	\label{eq:energy_weighted_fim}
\end{equation}
This factorization separates the deterministic derivatives of the sensing channel from the random symbol-energy weights. The complete FIM depends linearly on the energy of each transmitted symbol, while the symbol phases cancel. The energy distribution across subcarriers and OFDM symbols is retained, so equal total frame energy alone does not imply equal conditional CRBs.

\subsection{Average CRB and Constant-Modulus Optimality}

Let $\widehat{\boldsymbol{\vartheta}}$ be an estimator of the complete target-parameter vector that is locally unbiased conditioned on the transmitted symbols. For a fixed transmitted frame with $\boldsymbol{J}(\boldsymbol{S})\succ\boldsymbol{0}$, the estimation error satisfies
\begin{equation}
	\mathbb{E}_{\boldsymbol{W}_s}\!\left[
	(\widehat{\boldsymbol{\vartheta}}-\boldsymbol{\vartheta})
	(\widehat{\boldsymbol{\vartheta}}-\boldsymbol{\vartheta})^T
	\,\middle|\,\boldsymbol{S}
	\right]
	\succeq \boldsymbol{J}^{-1}(\boldsymbol{S}).
	\label{eq:conditional_crb}
\end{equation}
The first $Q$ diagonal entries of $\boldsymbol{J}^{-1}(\boldsymbol{S})$ give the conditional CRBs for the individual delays, and the next $Q$ give those for the individual Doppler shifts. Summing the corresponding entries gives a lower bound on the sum of the $Q$ delay or Doppler MSEs. For $a\in\{\tau,\nu\}$, let $[\cdot]_{aa}$ select rows and columns $1,\ldots,Q$ for $a=\tau$, or $Q+1,\ldots,2Q$ for $a=\nu$. These blocks are extracted after inverting the complete FIM, so the bounds account for the coupling among all jointly estimated parameters, including the unknown complex gains.

Averaging the conditional bound over the transmitted symbols gives the Miller-Chang bound~\cite{xiong2023fundamental} for the total delay or Doppler MSE,
\begin{equation}
	\begin{aligned}
		\overline{\operatorname{CRB}}_a
		&\triangleq \mathbb{E}_{\boldsymbol{S}}\!\left[
		\operatorname{Tr}\!\left([\boldsymbol{J}^{-1}(\boldsymbol{S})]_{aa}\right)
		\right]\\
		&=\operatorname{Tr}\!\left(
		\left[\mathbb{E}_{\boldsymbol{S}}[\boldsymbol{J}^{-1}(\boldsymbol{S})]\right]_{aa}
		\right).
	\end{aligned}
	\label{eq:average_crb}
\end{equation}

\begin{proposition}
	\label{prop:jensen}
	The average conditional CRB matrix for the complete parameter vector satisfies
	\begin{equation}
		\mathbb E_{\boldsymbol S}\!\left[\boldsymbol J^{-1}(\boldsymbol S)\right]
		\succeq \overline{\boldsymbol J}^{-1},
		\label{eq:matrix_crb_bound}
	\end{equation}
	and the average conditional CRBs for delay and Doppler satisfy
	\begin{equation}
		\overline{\operatorname{CRB}}_a
		\geq \operatorname{Tr}\!\left([\overline{\boldsymbol J}^{-1}]_{aa}\right),
		\quad a\in\{\tau,\nu\},
		\label{eq:jensen_trace}
	\end{equation}
	where $\overline{\boldsymbol J}\triangleq\mathbb E_{\boldsymbol S}[\boldsymbol J(\boldsymbol S)]$ is the average FIM. The right-hand sides of~\eqref{eq:matrix_crb_bound} and~\eqref{eq:jensen_trace} give the matrix and scalar Jensen-type bounds, respectively. Constant-modulus constellations such as PSK attain both bounds exactly for any finite $M$ and $N$.
\end{proposition}

\begin{proof}
	Since matrix inversion is operator convex on the positive-definite cone, Jensen's inequality gives
	\begin{equation*}
		\mathbb E_{\boldsymbol S}\!\left[\boldsymbol J^{-1}(\boldsymbol S)\right]
		\succeq \left(\mathbb E_{\boldsymbol S}\!\left[\boldsymbol J(\boldsymbol S)\right]\right)^{-1}
		= \overline{\boldsymbol J}^{-1}.
	\end{equation*}
	Taking the trace of the corresponding principal submatrix gives~\eqref{eq:jensen_trace}. Averaging~\eqref{eq:energy_weighted_fim} over the transmitted symbols yields
	\begin{equation}
		\begin{aligned}
			\overline{\boldsymbol J}
			&= \mathbb E_{\boldsymbol S}\!\left[\boldsymbol J(\boldsymbol S)\right] = \frac{2}{\sigma_W^2}\operatorname{Re}\!\left\{
			\boldsymbol G_0^H
			\operatorname{diag}\!\left(\mathbb E_{\boldsymbol S}\!\left[|\operatorname{vec}(\boldsymbol S)|^2\right]\right)
			\boldsymbol G_0
			\right\}\\
			&= \frac{2E_s}{\sigma_W^2}\operatorname{Re}\!\left\{\boldsymbol G_0^H\boldsymbol G_0\right\}.
		\end{aligned}
		\label{eq:mean_fim}
	\end{equation}
	At the same average symbol energy $E_s$, PSK and QAM therefore have the same Jensen-type bounds. Under PSK, $|S[k,m]|^2=E_s$. Substituting these energies into~\eqref{eq:energy_weighted_fim} gives
	\begin{equation*}
		\boldsymbol J(\boldsymbol S)
		= \frac{2}{\sigma_W^2}\operatorname{Re}\!\left\{\boldsymbol G_0^H(E_s\boldsymbol I_{NM})\boldsymbol G_0\right\}
		= \overline{\boldsymbol J}.
	\end{equation*}
	Thus $\boldsymbol J^{-1}(\boldsymbol S)=\overline{\boldsymbol J}^{-1}$ in every PSK frame, giving equality in~\eqref{eq:matrix_crb_bound} and~\eqref{eq:jensen_trace}.
	
	For nonconstant-modulus QAM, let $\boldsymbol J=\boldsymbol J(\boldsymbol S)$ and $\boldsymbol\Delta=\boldsymbol J-\overline{\boldsymbol J}$. The matrix inverse identity gives
	\begin{equation*}
		\begin{aligned}
			\boldsymbol J^{-1}-\overline{\boldsymbol J}^{-1}
			&= -\overline{\boldsymbol J}^{-1}\boldsymbol\Delta\overline{\boldsymbol J}^{-1}  + \overline{\boldsymbol J}^{-1}\boldsymbol\Delta\boldsymbol J^{-1}\boldsymbol\Delta\overline{\boldsymbol J}^{-1}.
		\end{aligned}
	\end{equation*}
	Taking the expectation over $\boldsymbol S$ and using $\mathbb E_{\boldsymbol S}[\boldsymbol\Delta]=\boldsymbol 0$ gives
	\begin{equation*}
		\mathbb E_{\boldsymbol S}[\boldsymbol J^{-1}]-\overline{\boldsymbol J}^{-1}
		= \overline{\boldsymbol J}^{-1}
		\mathbb E_{\boldsymbol S}\!\left[\boldsymbol\Delta\boldsymbol J^{-1}\boldsymbol\Delta\right]
		\overline{\boldsymbol J}^{-1}.
	\end{equation*}
	For any nonzero real vector $\boldsymbol x$, \eqref{eq:energy_weighted_fim} expresses $\boldsymbol x^T\boldsymbol\Delta\boldsymbol x$ as a weighted sum of independent centered symbol energies, with weights proportional to $|[\boldsymbol G_0\boldsymbol x]_i|^2$. At least one weight is positive because $\overline{\boldsymbol J}\succ\boldsymbol 0$ in~\eqref{eq:mean_fim}. The sum therefore has positive variance, so $\boldsymbol\Delta\boldsymbol x\ne\boldsymbol 0$ with positive probability. It follows that $\mathbb E_{\boldsymbol S}[\|\boldsymbol J^{-1/2}\boldsymbol\Delta\boldsymbol x\|_2^2]>0$. Thus the matrix gap is positive definite, and both average conditional CRBs strictly exceed their Jensen-type bounds.
\end{proof}

Proposition~\ref{prop:jensen} shows that, at a fixed average symbol energy, the conditional CRBs for delay and Doppler under PSK equal their respective Jensen-type bounds in every frame, for any finite $M$ and $N$. The minimum attained by PSK provides the benchmark for quantifying the CRB gap under QAM in Section~\ref{subsec:qam_psk_crb_gap}.

\subsection{Single-Target Closed-Form CRBs}
\label{subsec:single_target_crb}

To derive the single-target CRBs, we first state the following lemma, which gives a projection form of the equivalent FIM.

\begin{lemma}
	\label{lem:projected_fim}
	If $\boldsymbol{J}(\boldsymbol{S})\succ\boldsymbol{0}$, the joint delay–Doppler CRB matrix satisfies $[\boldsymbol{J}^{-1}(\boldsymbol{S})]_{\theta\theta}=\boldsymbol{J}_{\mathrm e}^{-1}(\boldsymbol{S})$, where the equivalent FIM is
	\begin{equation}
		\boldsymbol{J}_{\mathrm e}(\boldsymbol{S})
		= \frac{2}{\sigma_W^2}\operatorname{Re}\!\left\{\boldsymbol{D}^H\boldsymbol{\Pi}_{\boldsymbol{B}}^{\perp}\boldsymbol{D}\right\}.
		\label{eq:projected_fim}
	\end{equation}
	where
	\begin{equation}
		\boldsymbol{\Pi}_{\boldsymbol{B}}^{\perp}
		= \boldsymbol{I}_{NM}-\boldsymbol{B}(\boldsymbol{B}^H\boldsymbol{B})^{-1}\boldsymbol{B}^H
		\label{eq:gain_projector}
	\end{equation}
	is the orthogonal projector onto the null space of $\boldsymbol{B}^H$.
\end{lemma}

\begin{proof}
	See Appendix~\ref{app:schur}.
\end{proof}

For a single target with an unknown nonzero complex gain $\alpha$, the conditional range and velocity CRBs are
\begin{equation}
	\begin{aligned}
		\operatorname{CRB}(R\mid\boldsymbol{S})
		&=\frac{\sigma_W^2N^2\Delta R^2}{8\pi^2|\alpha|^2}\frac{S_{mm}}{\mathcal{D}},\\
		\operatorname{CRB}(v\mid\boldsymbol{S})
		&=\frac{\sigma_W^2M^2\Delta v^2}{8\pi^2|\alpha|^2}\frac{S_{kk}}{\mathcal{D}},
	\end{aligned}
	\label{eq:single_range_velocity_crbs}
\end{equation}
where $\mathcal{D}=S_{kk}S_{mm}-S_{km}^2>0$. With $p_{k,m}=|S[k,m]|^2$ and all sums taken over the OFDM frame, the energy-weighted central moments are
\begin{equation}
	\begin{aligned}
		S_{kk}&=\sum_{k,m}p_{k,m}(k-\bar{k})^2,\\
		S_{mm}&=\sum_{k,m}p_{k,m}(m-\bar{m})^2,\\
		S_{km}&=\sum_{k,m}p_{k,m}(k-\bar{k})(m-\bar{m}),
	\end{aligned}
	\label{eq:centered_moments}
\end{equation}
with the energy-weighted centers given by
\begin{equation}
	\bar{k}=\frac{\sum_{k,m}p_{k,m}k}{\sum_{k,m}p_{k,m}},
	\quad
	\bar{m}=\frac{\sum_{k,m}p_{k,m}m}{\sum_{k,m}p_{k,m}}.
	\label{eq:energy_weighted_centers}
\end{equation}
The moments $S_{kk}$ and $S_{mm}$ describe the energy-weighted spreads across subcarriers and OFDM symbols, respectively. The cross moment $S_{km}$ captures the coupling between delay and Doppler estimation. For fixed $S_{kk}$ and $S_{mm}$, a nonzero $S_{km}$ reduces $\mathcal{D}$ and increases both bounds.

Under constant-modulus signaling, $S_{km}=0$, so delay and Doppler decouple in the equivalent FIM. Substituting the constant symbol energies into~\eqref{eq:single_range_velocity_crbs} gives the following closed forms.

\begin{corollary}
	\label{cor:single_target_crbs}
	For a single target with $N,M\geq 2$ and $\alpha\neq0$, constant-modulus signaling attains the minimum average conditional range and velocity CRBs. The conditional bounds are the same in every frame and are given by
	\begin{equation}
		\begin{aligned}
			\operatorname{CRB}(R\mid\boldsymbol{S})
			&=\frac{3c^2}{8\pi^2\rho NM(N^2-1)\Delta f^2},\\
			\operatorname{CRB}(v\mid\boldsymbol{S})
			&=\frac{3\lambda^2}{8\pi^2\rho NM(M^2-1)T_{\mathrm{sym}}^2},
		\end{aligned}
		\label{eq:constant_range_velocity_crbs}
	\end{equation}
	where $\rho=|\alpha|^2E_s/\sigma_W^2$ is the received SNR per subcarrier and OFDM symbol.
\end{corollary}
\begin{proof}
	See Appendix~\ref{app:single} for the derivations of~\eqref{eq:single_range_velocity_crbs} and~\eqref{eq:constant_range_velocity_crbs}.
\end{proof}
By Proposition~\ref{prop:jensen}, constant-modulus signaling minimizes the average conditional delay and Doppler CRBs. The same optimality holds for the range and velocity CRBs, which differ only by the fixed scaling factors $\Delta R^2$ and $\Delta v^2$, respectively. 

\subsection{Relative CRB Gap Between QAM and PSK}
\label{subsec:qam_psk_crb_gap}

Proposition~\ref{prop:jensen} identifies the average conditional CRB attained by PSK. To quantify the increase under nonconstant-modulus QAM, we define the relative gap for $a\in\{\tau,\nu\}$ as
\begin{equation}
	\Gamma_a
	\triangleq
	\frac{\overline{\operatorname{CRB}}_{a,\mathrm{QAM}}-\overline{\operatorname{CRB}}_{a,\mathrm{PSK}}}
	{\overline{\operatorname{CRB}}_{a,\mathrm{PSK}}}.
	\label{eq:gap_relative_crb}
\end{equation}

Let $w_{\min}$ and $w_{\max}$ denote the minimum and maximum possible values of the normalized QAM symbol energy $|S[k,m]|^2/E_s$, and let $\alpha_{\max}=\max_q|\alpha_q|$. The normalized mean FIM is
\begin{equation}
	\boldsymbol{J}_0
	\triangleq \frac{\sigma_W^2}{2E_sNM}\overline{\boldsymbol{J}}
	=\frac{1}{NM}\operatorname{Re}\!\left\{\boldsymbol{G}_0^H\boldsymbol{G}_0\right\}.
	\label{eq:gap_normalized_mean_fim}
\end{equation}

\begin{theorem}[Relative CRB gap between QAM and PSK]
	\label{thm:qam_psk_crb_gap}
	Consider PSK and nonconstant-modulus QAM under the common symbol-energy normalization in Section~\ref{sec:continuous}. Suppose that $w_{\min}>0$ and $\boldsymbol{J}_0\succ\boldsymbol{0}$. For $a\in\{\tau,\nu\}$, the relative gap satisfies
	\begin{equation}
		\begin{aligned}
		 \frac{\kappa-1}{w_{\max}\lambda_{\max}(\boldsymbol{J}_0)}\frac{Q}{NM} \leq \Gamma_a  \leq \frac{(\kappa-1)(2+8\pi^2\alpha_{\max}^2)}
			{w_{\min}\lambda_{\min}(\boldsymbol{J}_0)}\frac{Q}{NM}.
		\end{aligned}
		\label{eq:gap_finite_frame_bounds}
	\end{equation}
	For a fixed QAM symbol distribution, if $\alpha_{\max}$ is uniformly bounded and the eigenvalues of $\boldsymbol{J}_0$ are uniformly bounded above and away from zero, then
	\begin{equation}
		\Gamma_\tau=\Theta\!\left(\frac{Q}{NM}\right),
		\quad
		\Gamma_\nu=\Theta\!\left(\frac{Q}{NM}\right).
		\label{eq:gap_asymptotic_order}
	\end{equation}
\end{theorem}
\begin{proof}
	The key to the proof is to first use a matrix inverse identity to bound the CRB increment in terms of the second moment of the normalized FIM perturbation and then exploit the structure of the multi-target channel Jacobian to derive upper and lower bounds on $\Gamma_a$ with a common factor $Q/(NM)$. For a detailed proof, see Appendix~\ref{app:qam_psk_crb_gap}.
\end{proof}
The factors $\Delta R^2$ and $\Delta v^2$ cancel in their respective ratios. Thus, $\Gamma_\tau$ and $\Gamma_\nu$ also give the relative range and velocity CRB gaps.

\section{Maximum-Likelihood Estimation of Delay and Doppler}
\label{sec:fft_ml}

Dividing the received subcarrier samples by the known nonzero communication symbols gives $Z[k,m]=Y_s[k,m]/S[k,m]$. The resulting noise samples $\eta[k,m]=W_s[k,m]/S[k,m]$ remain conditionally independent and satisfy
\begin{equation}
	\eta[k,m]\mid\boldsymbol{S}
	\sim
	\mathcal{CN}\!\left(0,\frac{\sigma_W^2}{p_{k,m}}\right),
	\quad p_{k,m}=|S[k,m]|^2.
	\label{eq:divider_noise}
\end{equation}
The 2D FFT estimator applies an $N$-point IFFT across subcarriers and an $M$-point FFT along slow time, yielding $\boldsymbol{V}=\boldsymbol{F}_N^H\boldsymbol{Z}\boldsymbol{F}_M$. The $Q$ dominant distinct local maxima of $|V[r,s]|^2$ in the admissible search region give the delay and Doppler estimates on the FFT grid. This grid restriction introduces quantization errors for off-grid targets. We use continuous-parameter ML estimation to assess estimation accuracy against the derived CRBs.

Using the vectorized observation model~\eqref{eq:linear_observation}, the joint ML estimate for a known number of targets solves
\begin{equation}
	(\widehat{\boldsymbol{\theta}},\widehat{\boldsymbol{\alpha}})
	=
	\underset{\boldsymbol{\theta},\boldsymbol{\alpha}}{\arg\min}\;
	\left\|
	\boldsymbol{y}-\boldsymbol{B}(\boldsymbol{\theta})\boldsymbol{\alpha}
	\right\|_2^2.
	\label{eq:joint_ml}
\end{equation}
For a fixed $\boldsymbol{\theta}$ at which $\boldsymbol{B}(\boldsymbol{\theta})$ has full column rank, minimizing~\eqref{eq:joint_ml} over the complex gains gives the least-squares estimate
\begin{equation}
	\widehat{\boldsymbol{\alpha}}(\boldsymbol{\theta})
	=
	\left[
	\boldsymbol{B}^H(\boldsymbol{\theta})\boldsymbol{B}(\boldsymbol{\theta})
	\right]^{-1}
	\boldsymbol{B}^H(\boldsymbol{\theta})\boldsymbol{y}.
	\label{eq:conditional_ml_gains}
\end{equation}
Substituting~\eqref{eq:conditional_ml_gains} into~\eqref{eq:joint_ml} eliminates the gain vector and yields the concentrated ML criterion
\begin{equation}
	\begin{aligned}
		\widehat{\boldsymbol{\theta}}
		&=
		\underset{\boldsymbol{\theta}}{\arg\min}\;
		\boldsymbol{y}^H\boldsymbol{\Pi}_{\boldsymbol{B}(\boldsymbol{\theta})}^{\perp}\boldsymbol{y}\\
		&=
		\underset{\boldsymbol{\theta}}{\arg\max}\;
		\boldsymbol{y}^H\boldsymbol{B}(\boldsymbol{\theta})
		\left[
		\boldsymbol{B}^H(\boldsymbol{\theta})\boldsymbol{B}(\boldsymbol{\theta})
		\right]^{-1}
		\boldsymbol{B}^H(\boldsymbol{\theta})\boldsymbol{y},
	\end{aligned}
	\label{eq:concentrated_ml}
\end{equation}
where the projector in~\eqref{eq:gain_projector} is evaluated at the candidate parameters. The joint gain estimate is then $\widehat{\boldsymbol{\alpha}}=\widehat{\boldsymbol{\alpha}}(\widehat{\boldsymbol{\theta}})$.

For a single target, \eqref{eq:concentrated_ml} reduces to
\begin{equation}
	(\widehat{\tau},\widehat{\nu})
	=
	\underset{(\tau,\nu)\in\Omega}{\arg\max}\;
	\frac{
		\left|
		\displaystyle\sum_{k,m}
		S^*[k,m]Y_s[k,m]\,
		e^{j2\pi k\tau/N}e^{-j2\pi m\nu/M}
		\right|^2
	}{
		\displaystyle\sum_{k,m}|S[k,m]|^2
	},
	\label{eq:single_target_ml}
\end{equation}
where $\Omega$ is the admissible continuous delay–Doppler search region. Since $S^*[k,m]Y_s[k,m]=p_{k,m}Z[k,m]$, \eqref{eq:single_target_ml} weights the division output by the symbol energies, accounting for the unequal conditional noise variances in~\eqref{eq:divider_noise}.

Under constant-modulus transmission, $p_{k,m}=E_s$, and the objective in~\eqref{eq:single_target_ml} evaluated at an FFT grid point $(\tau,\nu)=(r,s)$ equals $E_s|V[r,s]|^2$. Thus, for a single target under constant-modulus signaling, ML estimation is equivalent to 2D FFT peak search without the grid constraint. In the simulations, the FFT peaks initialize local continuous optimization of~\eqref{eq:concentrated_ml} for multiple targets or~\eqref{eq:single_target_ml} for a single target.

\begin{figure*}[t]
	\centering
	\subfigure[FFT ]{ \label{locate}
		\begin{minipage}[t]{0.23\linewidth}
			\centering
			\includegraphics[width=1\textwidth]{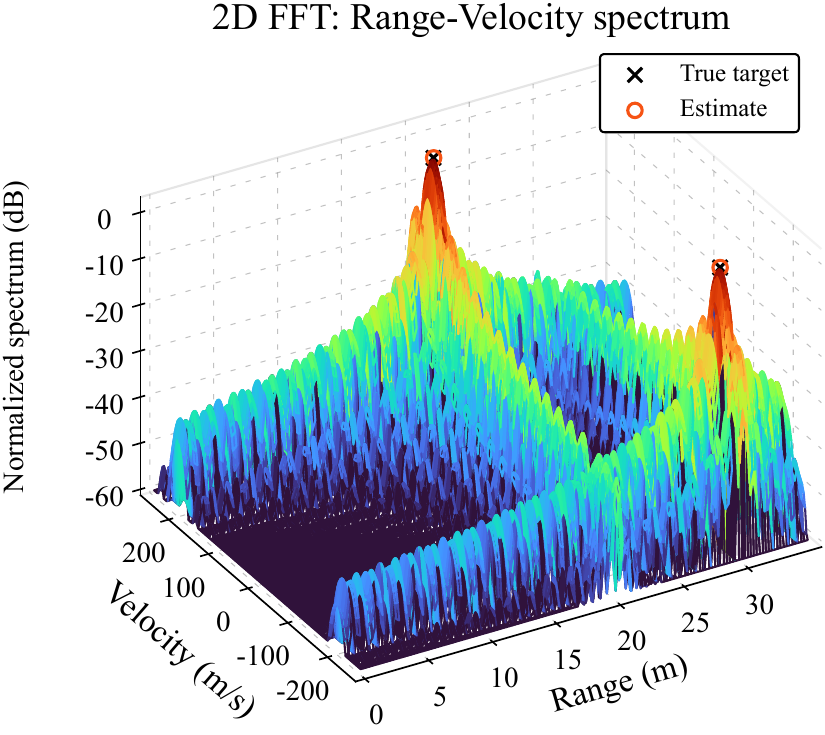}  
		\end{minipage}
	}%
	\subfigure[FFT, top view]{ \label{ber_2user}
		\begin{minipage}[t]{0.23\linewidth}
			\centering
			\includegraphics[width=1\textwidth]{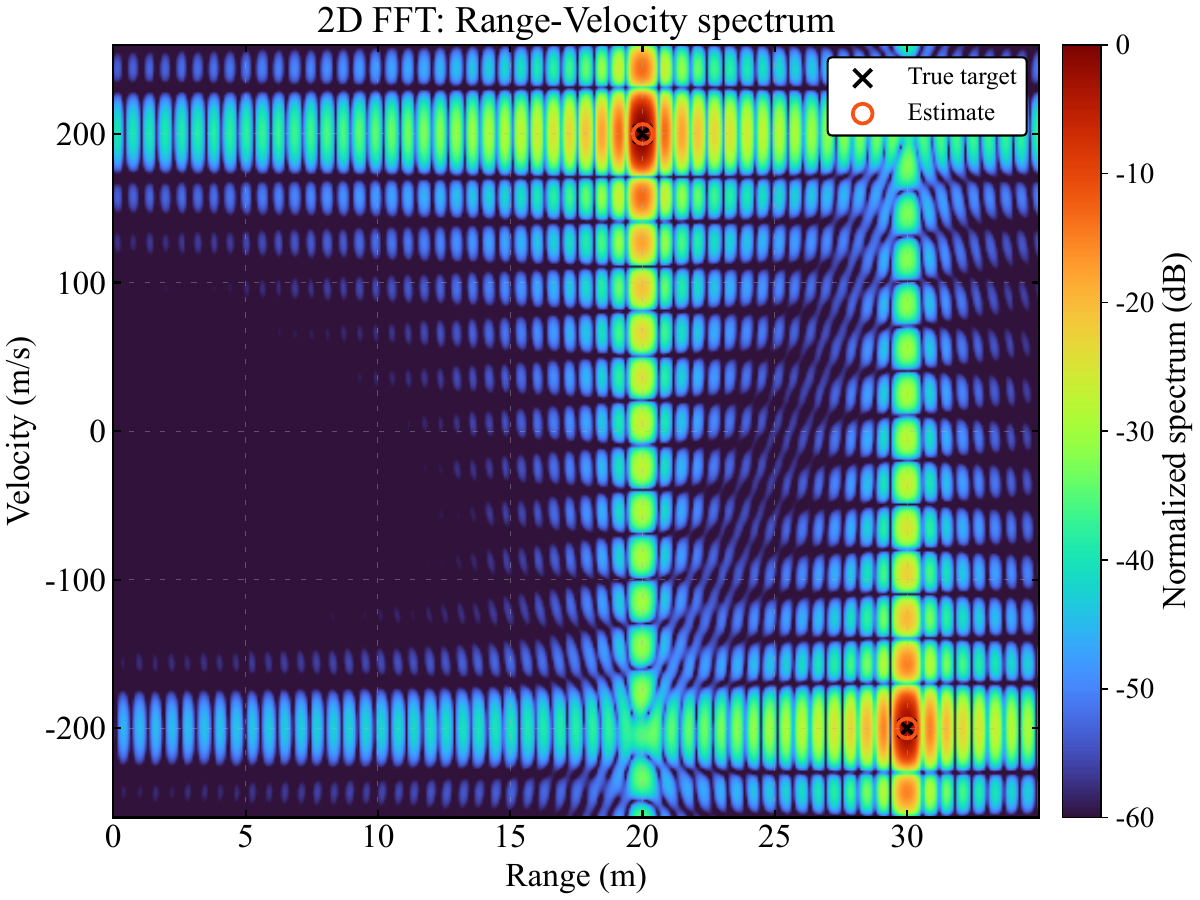} 
		\end{minipage}
	}
	\subfigure[MUSIC ]{ \label{ber_4user}
		\begin{minipage}[t]{0.23\linewidth}
			\centering
			\includegraphics[width=1\textwidth]{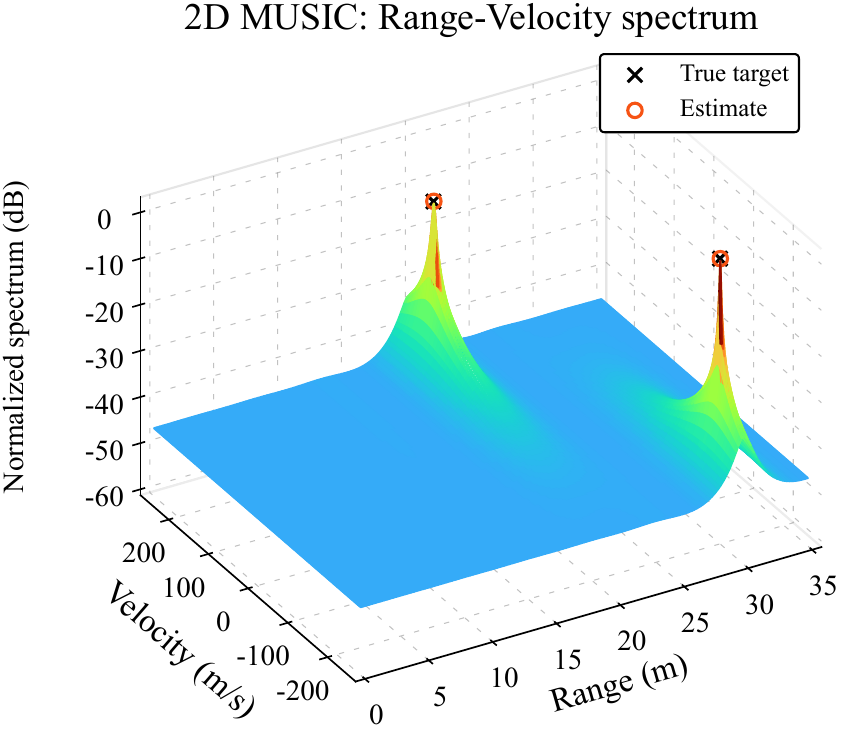} 
		\end{minipage}
	}
	\subfigure[MUSIC, top view]{ \label{ber_6user}
		\begin{minipage}[t]{0.23\linewidth}
			\centering
			\includegraphics[width=1\textwidth]{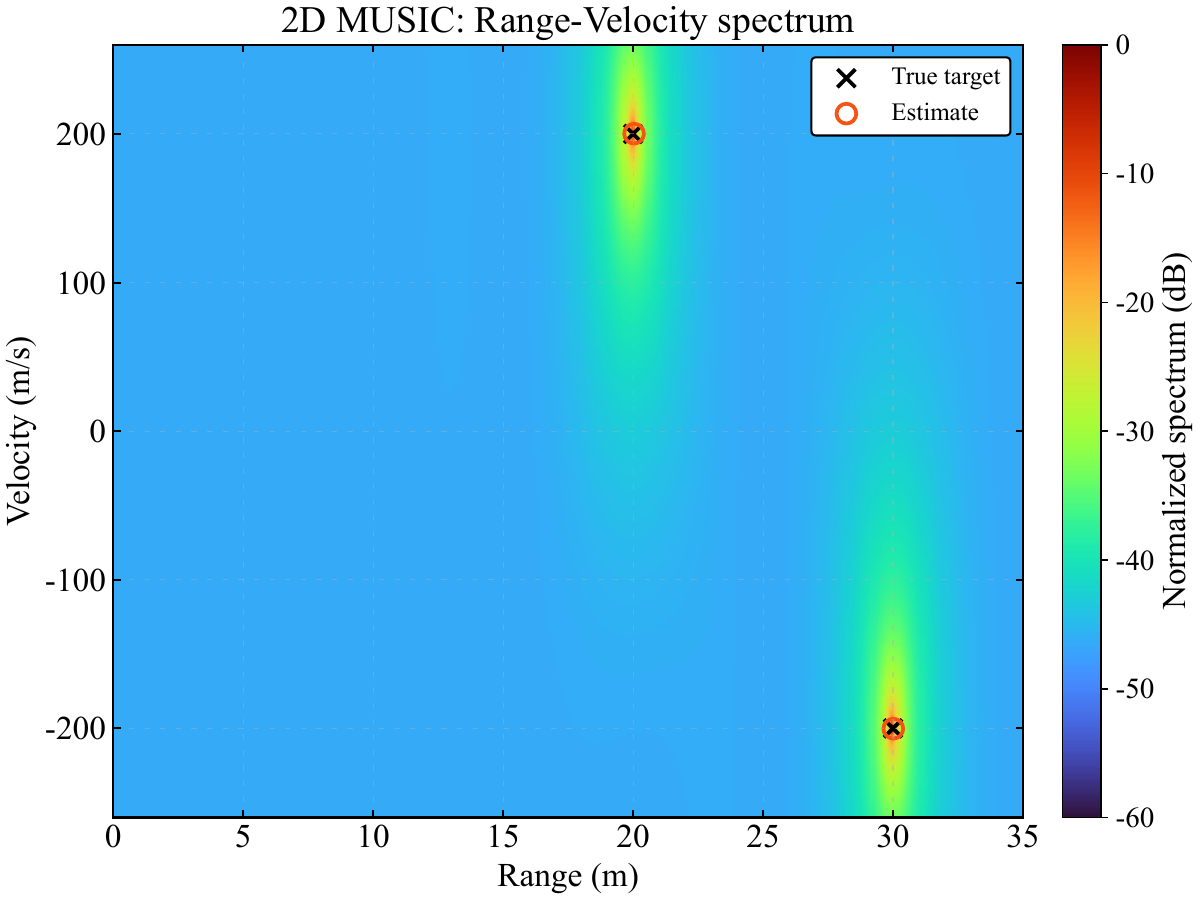} 
		\end{minipage}
	}
	\centering
	\caption{Range-velocity spectra of 2D FFT and MUSIC under 16-QAM transmission with $N=256$, $M=16$, and $Q=2$. (a) FFT spectrum. (b) FFT spectrum, top view. (c) MUSIC spectrum. (d) MUSIC spectrum, top view.}
	\label{fig:fft_music_spectra}
\end{figure*}

\begin{figure*}[t]
	\centering
	\subfigure[$N = 16, M = 4, Q = 12$]{  
		\begin{minipage}[t]{0.23\linewidth}
			\centering
			\includegraphics[width=1\textwidth]{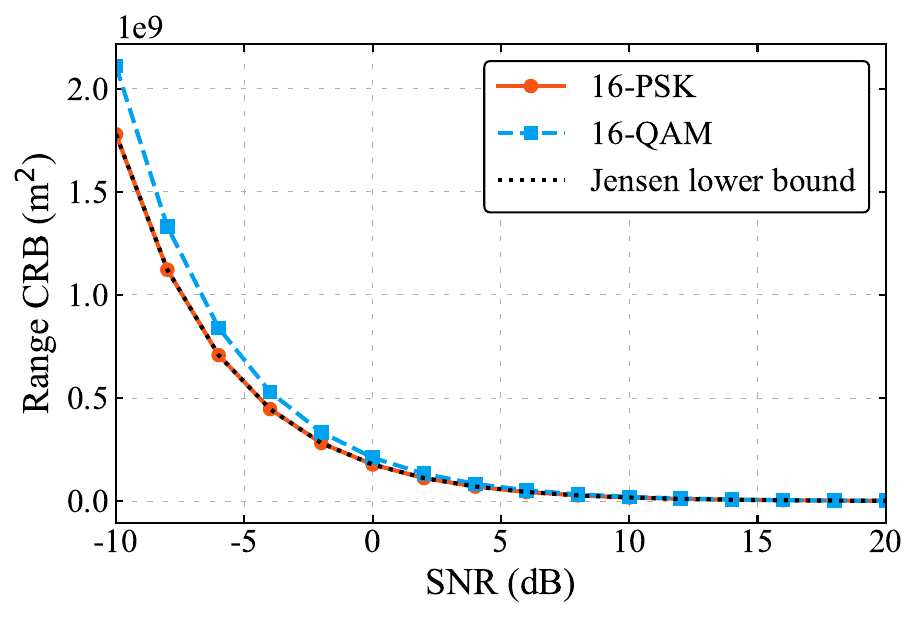}  
		\end{minipage}
	}%
	\subfigure[$N = 16, M = 4, Q = 12$]{ 
		\begin{minipage}[t]{0.23\linewidth}
			\centering
			\includegraphics[width=1\textwidth]{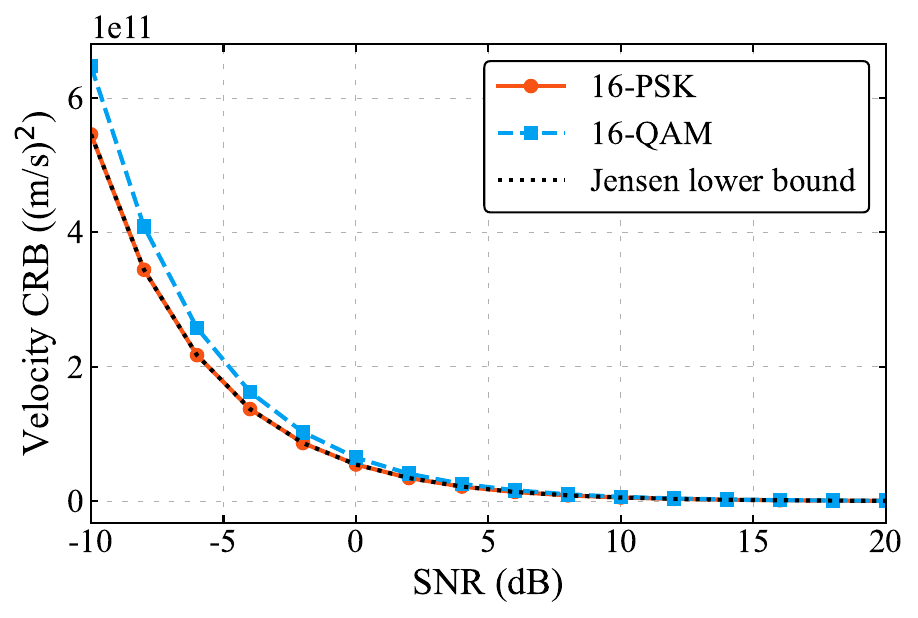} 
		\end{minipage}
	}
	\subfigure[$N = 128, M = 64, Q = 60$]{ 
		\begin{minipage}[t]{0.23\linewidth}
			\centering
			\includegraphics[width=1\textwidth]{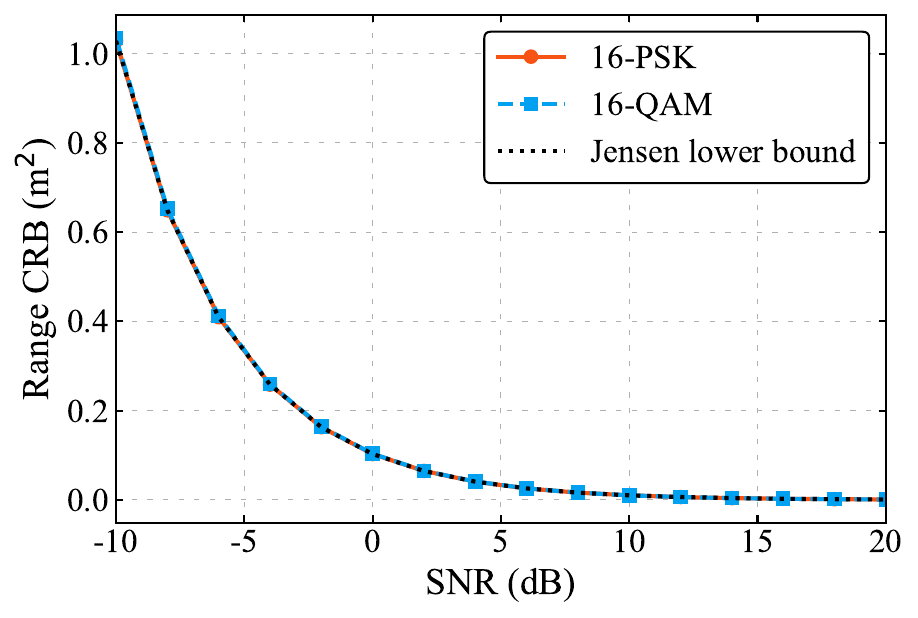} 
		\end{minipage}
	}
	\subfigure[$N = 128, M = 64, Q = 60$]{ 
		\begin{minipage}[t]{0.23\linewidth}
			\centering
			\includegraphics[width=1\textwidth]{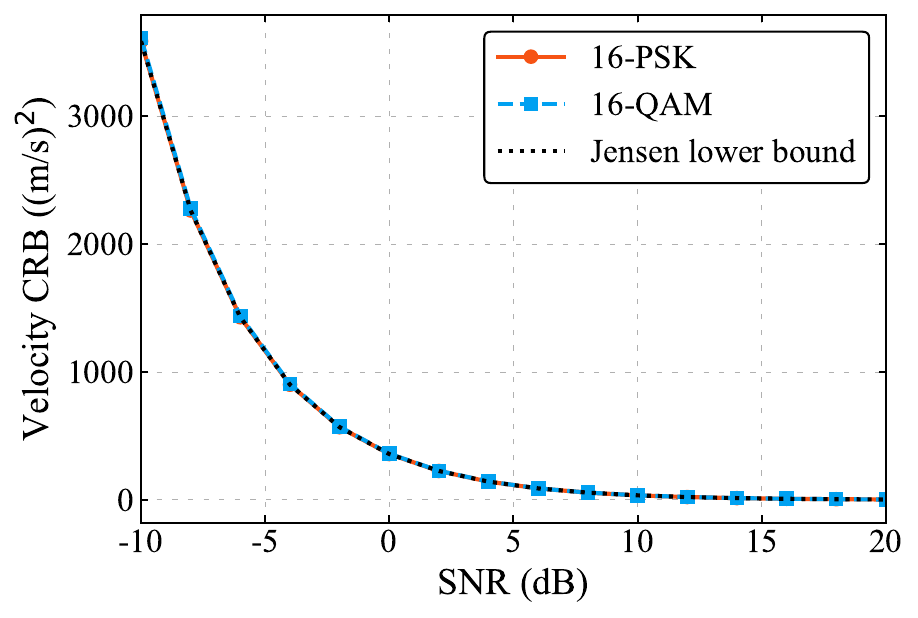} 
		\end{minipage}
	}
	\caption{Average range and velocity CRBs of 16-PSK and 16-QAM compared with the Jensen lower bound. (a)(b) $N=16$, $M=4$, and $Q=12$. (c)(d) $N=128$, $M=64$, and $Q=60$.}
	\label{fig:average_crbs}
\end{figure*}

\begin{figure}[t]
	\centering
	\subfigure[Range CRB gap]{  
		\begin{minipage}[t]{0.5\linewidth}
			\centering
			\includegraphics[width=1\textwidth]{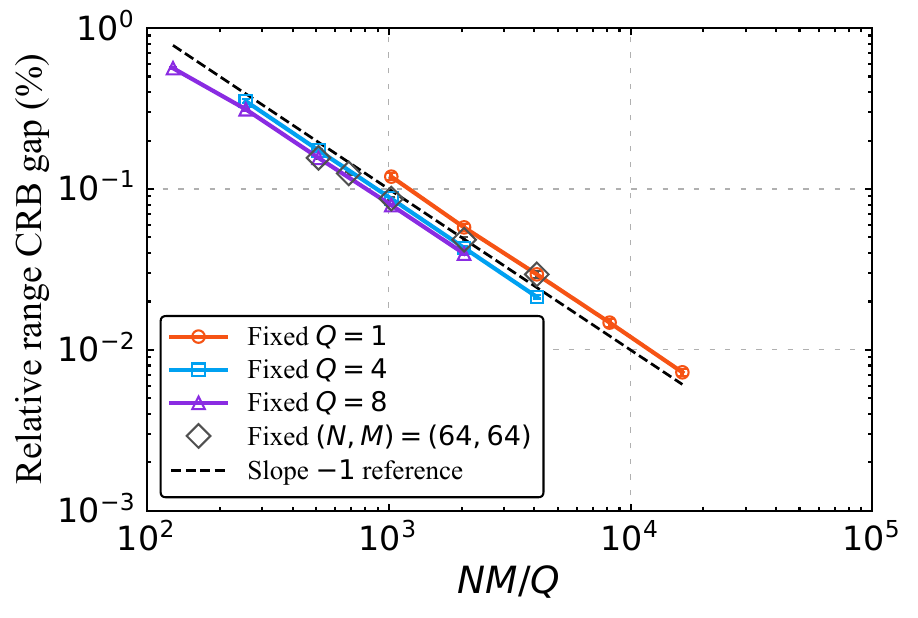}  
		\end{minipage}
	}%
	\subfigure[Velocity CRB gap]{ 
		\begin{minipage}[t]{0.5\linewidth}
			\centering
			\includegraphics[width=1\textwidth]{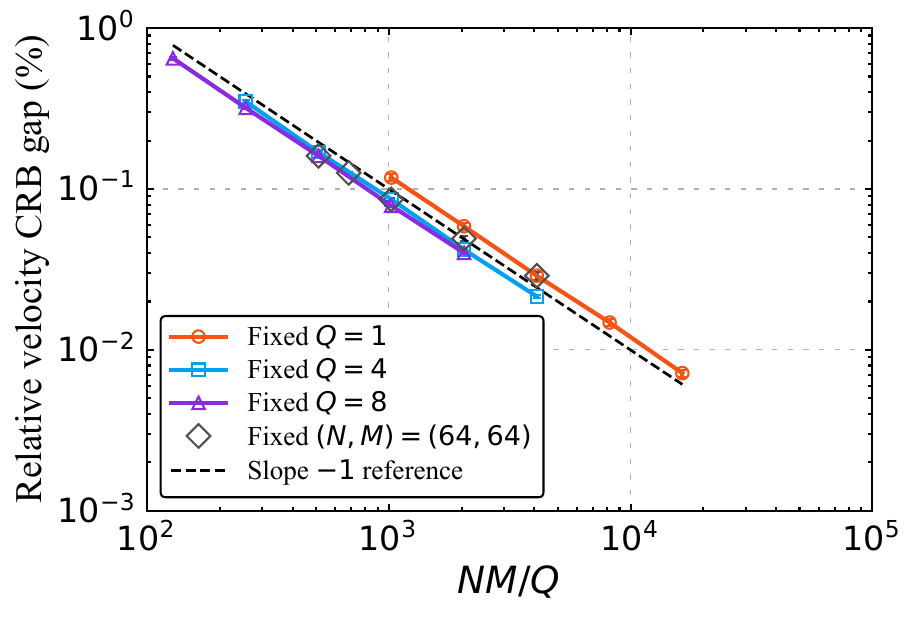} 
		\end{minipage}
	}
	\caption{Relative CRB gap between 16-PSK and 16-QAM versus $NM/Q$ for (a) range estimation and (b) velocity estimation.}
	\label{fig:relative_crb_gap}
\end{figure}

\begin{figure}[t]
	\centering
	\subfigure[Range RMSE vs CRB]{  
		\begin{minipage}[t]{0.5\linewidth}
			\centering
			\includegraphics[width=1\textwidth]{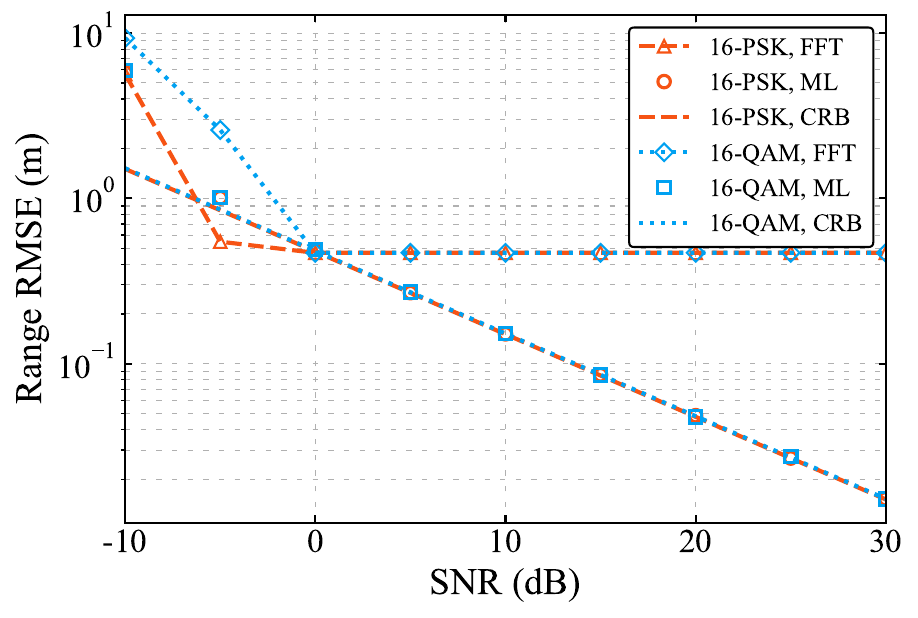}  
		\end{minipage}
	}%
	\subfigure[Velocity RMSE vs CRB]{ 
		\begin{minipage}[t]{0.5\linewidth}
			\centering
			\includegraphics[width=1\textwidth]{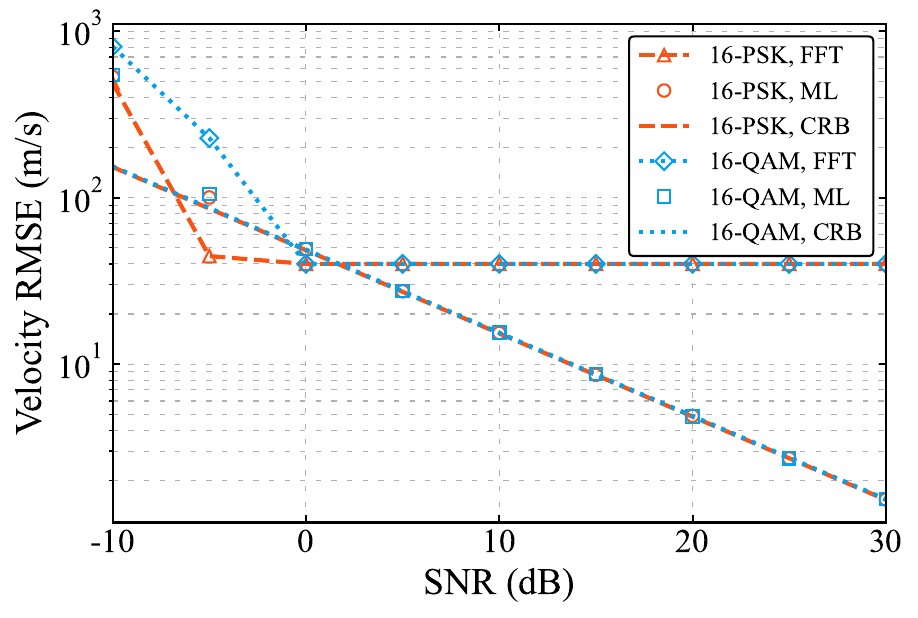} 
		\end{minipage}
	}
	\caption{Single-target~($Q=1$) FFT and ML estimation performance compared with the square roots of the average CRBs under 16-PSK and 16-QAM transmission, with $N=16$ and $M=4$. (a) Range RMSE. (b) Velocity RMSE.}
	\label{fig:single_target_estimation}
\end{figure}

\begin{figure}[t]
	\centering
	\subfigure[Range RMSE vs CRB]{  
		\begin{minipage}[t]{0.5\linewidth}
			\centering
			\includegraphics[width=1\textwidth]{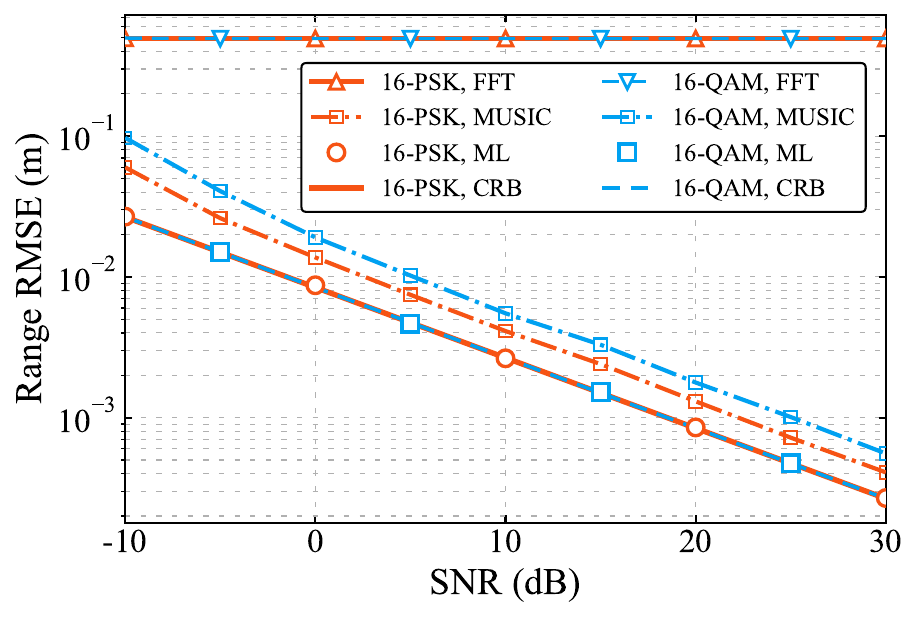}  
		\end{minipage}
	}%
	\subfigure[Velocity RMSE vs CRB]{ 
		\begin{minipage}[t]{0.5\linewidth}
			\centering
			\includegraphics[width=1\textwidth]{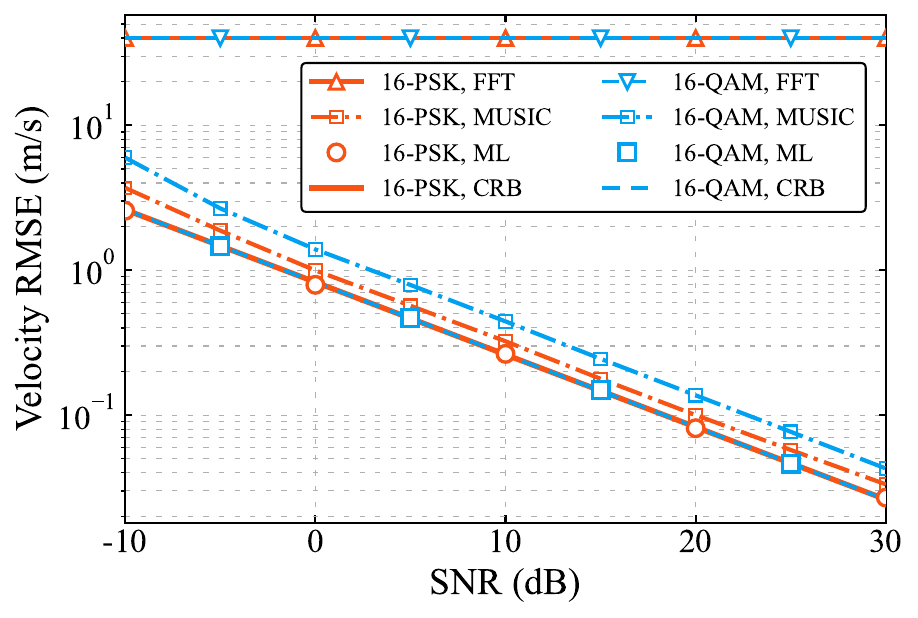} 
		\end{minipage}
	}
	\caption{Two-target~($Q=2$) FFT, MUSIC, and ML estimation performance compared with the square roots of the average CRBs under 16-PSK and 16-QAM transmission, with $N=256$ and $M=16$. (a) Range RMSE. (b) Velocity RMSE.}
	\label{fig:two_target_estimation}
\end{figure}

\section{Numerical Results}
\label{sec:numerical}

We evaluate the constellation-dependent CRBs and compare them with the estimation accuracy of FFT, ML, and MUSIC. The carrier frequency and subcarrier spacing are set to $f_c=30$~GHz and $\Delta f=960$~kHz, respectively, with a CP length of $N/4$. Unless otherwise specified, we consider independently generated 16-PSK and 16-QAM symbols normalized to the same average energy $E_s=1$, and define the SNR as $E_s/\sigma_W^2$.

The average conditional CRBs are evaluated over the transmitted communication data. All Monte Carlo simulations are performed with $10^4$ trials. The frame dimensions and number of targets are specified for each experiment.

For the MUSIC baseline~\cite{music1986}, we form snapshots by vectorizing overlapping $N_s\times M_s$ subarrays of $\boldsymbol{Z}$. For each subarray coordinate, we divide the snapshot entries and the corresponding steering-vector entry by the square root of the mean of $|S[k,m]|^{-2}$ over the sliding positions. This normalization makes the average conditional noise covariance proportional to the identity. The $Q$ dominant distinct peaks of the resulting 2D MUSIC pseudospectrum give the paired delay and Doppler estimates.

\subsection{Two-Dimensional FFT and MUSIC Spectra}

Fig.~\ref{fig:fft_music_spectra} presents the range-velocity spectra obtained by 2D FFT and MUSIC under 16-QAM transmission, with $N=256$, $M=16$, and $Q=2$. The two targets are located at 20~m and 30~m, with radial velocities of 200~m/s and $-200$~m/s, respectively. Both methods produce distinct peaks near the true target positions, demonstrating joint range-velocity estimation under the adopted observation model. Compared with FFT, MUSIC produces more concentrated spectral peaks. The corresponding top views further illustrate the agreement between the estimated and true target locations.

\subsection{Average CRBs and Constant-Modulus Optimality}

Fig.~\ref{fig:average_crbs} compares the average range and velocity CRBs of 16-PSK and 16-QAM with the Jensen lower bound. Two configurations are considered: $N=16$, $M=4$, $Q=12$, and $N=128$, $M=64$, $Q=60$. In both configurations, the PSK curves coincide with the Jensen lower bound, confirming equality in~\eqref{eq:jensen_trace} under constant-modulus transmission, as established in Proposition~\ref{prop:jensen}.

However, the CRB advantage of PSK over QAM is not always pronounced. As shown in Fig.~\ref{fig:average_crbs}(c) and Fig.~\ref{fig:average_crbs}(d), the PSK–QAM CRB gap becomes negligible in the configuration with larger numbers of subcarriers and OFDM symbols, despite the increased number of targets. This observation suggests that large OFDM frames can substantially reduce the influence of the communication constellation on the range and velocity CRBs.

\subsection{Scaling of the Relative CRB Gap}

Fig.~\ref{fig:relative_crb_gap} compares the simulated relative CRB gaps with the $Q/(NM)$ scaling predicted by Theorem~\ref{thm:qam_psk_crb_gap}. The relative gaps for both range and velocity decrease approximately in proportion to $Q/(NM)$ when the frame dimensions vary at fixed $Q$. Varying $Q$ at fixed frame dimensions produces the same trend. At the larger values of $NM/Q$ considered here, 16-QAM and 16-PSK yield nearly identical CRBs for both range and velocity estimation.

\subsection{Single-Target Estimation Performance}

Fig.~\ref{fig:single_target_estimation} compares the range and velocity RMSEs of FFT and continuous ML estimation with the corresponding CRBs for $N=16$, $M=4$, and $Q=1$. The PSK bounds are obtained directly from~\eqref{eq:constant_range_velocity_crbs}, whereas the QAM bounds are calculated by averaging the conditional expressions in~\eqref{eq:single_range_velocity_crbs} over the transmitted data. The PSK and QAM CRBs nearly coincide, indicating little constellation dependence in this single-target setting.

As the SNR increases, the ML results closely approach the corresponding bounds, providing numerical validation of the single-target CRB expressions. FFT instead exhibits an error floor because its estimates are restricted to discrete delay--Doppler grid points. For the off-grid target considered here, increasing the SNR cannot eliminate this quantization error. Note that at some low-SNR points, the FFT RMSE falls below the root CRB. This behavior does not contradict~\eqref{eq:conditional_crb}, which assumes locally unbiased estimation, since FFT grid selection can produce biased estimates.

\subsection{Two-Target Estimation Performance}

Fig.~\ref{fig:two_target_estimation} extends the comparison to two targets with $N=256$ and $M=16$, including FFT, MUSIC, and joint ML estimation based on~\eqref{eq:concentrated_ml}. The PSK and QAM CRBs remain nearly indistinguishable. For both constellations, joint ML estimation closely approaches the theoretical bounds at sufficiently high SNR, supporting the multi-target CRB analysis in the presence of unknown complex gains.

MUSIC achieves lower RMSE than FFT, whose performance remains limited by the discrete estimation grid. Unlike the nearly overlapping CRBs, the MUSIC results exhibit a more visible constellation dependence, with PSK yielding lower errors than QAM. This difference is consistent with the nonuniform noise amplification introduced by symbolwise division under QAM transmission, as characterized by~\eqref{eq:divider_noise}. Thus, similar estimation CRBs do not necessarily imply similar performance for a particular sensing algorithm for different constellations.

\section{Conclusion}\label{sec:conclusion}
This paper investigated how communication constellations affect joint range–velocity estimation with OFDM signals. Starting from the continuous-time transmit and receive signals, we established an equivalent discrete sensing model and specified its validity conditions. Under joint estimation of target delays, Doppler shifts, and unknown complex gains, we proved that constant-modulus constellations minimize the average conditional CRBs by exactly attaining the Jensen lower bound.
The numerical results showed that this theoretical optimality does not always produce a pronounced performance advantage over QAM. The PSK–QAM CRB gap becomes negligible for the larger numbers of subcarriers and OFDM symbols considered. ML estimation closely approaches the corresponding CRBs under both constellations, confirming the validity of the theoretical analysis. FFT exhibits error floors imposed by its integer-grid constraint, whereas MUSIC achieves substantially lower estimation errors and retains a more visible constellation dependence than the CRBs. These findings show that the benefit of constellation selection depends on both the OFDM parameters and the estimation algorithm.

\appendices
\section{Projection Form of the Equivalent FIM}
\label{app:schur}

To evaluate the delay–Doppler block of the complete inverse FIM directly, we use its Schur-complement representation. Recall that $\boldsymbol{\vartheta}=[\boldsymbol{\theta}^T,\boldsymbol{\xi}^T]^T$, so the complete FIM can be partitioned as
\begin{equation}
	\boldsymbol{J}(\boldsymbol{S})
	= \begin{bmatrix}
		\boldsymbol{J}_{\theta\theta} & \boldsymbol{J}_{\theta\xi}\\
		\boldsymbol{J}_{\theta\xi}^T & \boldsymbol{J}_{\xi\xi}
	\end{bmatrix}.
	\label{eq:fim_partition}
\end{equation}
When the complete FIM is positive definite, the block-inversion identity gives
\begin{equation}
	[\boldsymbol{J}^{-1}(\boldsymbol{S})]_{\theta\theta}
	= \boldsymbol{J}_{\mathrm e}^{-1}(\boldsymbol{S}),
	\label{eq:inverse_fim_block_relation}
\end{equation}
where the equivalent FIM for the delays and Doppler is
\begin{equation}
	\boldsymbol{J}_{\mathrm e}(\boldsymbol{S})
	= \boldsymbol{J}_{\theta\theta}-\boldsymbol{J}_{\theta\xi}\boldsymbol{J}_{\xi\xi}^{-1}\boldsymbol{J}_{\theta\xi}^T.
	\label{eq:schur_fim}
\end{equation}
With the complete Jacobian $\boldsymbol{G}_\mu=[\boldsymbol{D},\boldsymbol{B},j\boldsymbol{B}]$, the FIM blocks associated with the delay–Doppler and gain parameters are
\begin{equation}
	\begin{aligned}
		\boldsymbol{J}_{\theta\theta}
		&= \frac{2}{\sigma_W^2}\operatorname{Re}\{\boldsymbol{D}^H\boldsymbol{D}\},\\
		\boldsymbol{J}_{\theta\xi}
		&= \frac{2}{\sigma_W^2}
		\left[\operatorname{Re}\{\boldsymbol{D}^H\boldsymbol{B}\},-\operatorname{Im}\{\boldsymbol{D}^H\boldsymbol{B}\}\right],\\
		\boldsymbol{J}_{\xi\xi}
		&= \frac{2}{\sigma_W^2}
		\begin{bmatrix}
			\operatorname{Re}\{\boldsymbol{B}^H\boldsymbol{B}\} & -\operatorname{Im}\{\boldsymbol{B}^H\boldsymbol{B}\}\\
			\operatorname{Im}\{\boldsymbol{B}^H\boldsymbol{B}\} & \operatorname{Re}\{\boldsymbol{B}^H\boldsymbol{B}\}
		\end{bmatrix}.
	\end{aligned}
	\label{eq:fim_blocks}
\end{equation}
Since $\boldsymbol{B}$ has full column rank, $\boldsymbol{B}^H\boldsymbol{B}$ is invertible. For any invertible complex matrix $\boldsymbol{R}$, its real block representation satisfies
\begin{equation}
	\begin{bmatrix}
		\operatorname{Re}\{\boldsymbol{R}\} & -\operatorname{Im}\{\boldsymbol{R}\}\\
		\operatorname{Im}\{\boldsymbol{R}\} & \operatorname{Re}\{\boldsymbol{R}\}
	\end{bmatrix}^{-1}
	=
	\begin{bmatrix}
		\operatorname{Re}\{\boldsymbol{R}^{-1}\} & -\operatorname{Im}\{\boldsymbol{R}^{-1}\}\\
		\operatorname{Im}\{\boldsymbol{R}^{-1}\} & \operatorname{Re}\{\boldsymbol{R}^{-1}\}
	\end{bmatrix}.
	\label{eq:real_inverse_identity}
\end{equation}
Applying this identity with $\boldsymbol{R}=\boldsymbol{B}^H\boldsymbol{B}$ gives
\begin{equation}
	\boldsymbol{J}_{\xi\xi}^{-1}
	= \frac{\sigma_W^2}{2}
	\begin{bmatrix}
		\operatorname{Re}\{(\boldsymbol{B}^H\boldsymbol{B})^{-1}\} & -\operatorname{Im}\{(\boldsymbol{B}^H\boldsymbol{B})^{-1}\}\\
		\operatorname{Im}\{(\boldsymbol{B}^H\boldsymbol{B})^{-1}\} & \operatorname{Re}\{(\boldsymbol{B}^H\boldsymbol{B})^{-1}\}
	\end{bmatrix}.
	\label{eq:gain_information_inverse}
\end{equation}
Thus, we further have
\begin{equation}
	\boldsymbol{J}_{\theta\xi}\boldsymbol{J}_{\xi\xi}^{-1}\boldsymbol{J}_{\theta\xi}^T
	= \frac{2}{\sigma_W^2}\operatorname{Re}\!\left\{\boldsymbol{D}^H\boldsymbol{B}(\boldsymbol{B}^H\boldsymbol{B})^{-1}\boldsymbol{B}^H\boldsymbol{D}\right\}.
	\label{eq:schur_removed_information}
\end{equation}
Substituting this expression and $\boldsymbol{J}_{\theta\theta}$ into the Schur complement gives
\begin{equation}
	\begin{aligned}
		\boldsymbol{J}_{\mathrm e}(\boldsymbol{S})
		&= \frac{2}{\sigma_W^2}\operatorname{Re}\!\left\{\boldsymbol{D}^H\left[\boldsymbol{I}_{NM}-\boldsymbol{B}(\boldsymbol{B}^H\boldsymbol{B})^{-1}\boldsymbol{B}^H\right]\boldsymbol{D}\right\}\\
		&= \frac{2}{\sigma_W^2}\operatorname{Re}\!\left\{\boldsymbol{D}^H\boldsymbol{\Pi}_{\boldsymbol{B}}^{\perp}\boldsymbol{D}\right\},
	\end{aligned}
	\label{eq:app_projected_fim}
\end{equation}
which completes the proof.

\section{Proof of Corollary~\ref{cor:single_target_crbs}}
\label{app:single}

To compute the single-target equivalent FIM in~\eqref{eq:projected_fim}, we expand the projector in~\eqref{eq:gain_projector} and evaluate the two products $\boldsymbol{D}^H\boldsymbol{D}$ and $\boldsymbol{B}^H\boldsymbol{D}$. For a single target, $\boldsymbol{B}$ has one column, and the delay–Doppler Jacobian becomes
\begin{equation}
	\boldsymbol{D}
	= j2\pi\alpha
	\left[
	-\frac{1}{N}\boldsymbol{K}\boldsymbol{B},
	\frac{1}{M}\boldsymbol{L}\boldsymbol{B}
	\right].
	\label{eq:app_single_jacobian}
\end{equation}
Using the energy-weighted centers in~\eqref{eq:energy_weighted_centers}, the gain–parameter coupling product is
\begin{equation}
	\begin{aligned}
		\boldsymbol{B}^H\boldsymbol{D}
		&= j2\pi\alpha
		\left[
		-\frac{\boldsymbol{B}^H\boldsymbol{K}\boldsymbol{B}}{N},
		\frac{\boldsymbol{B}^H\boldsymbol{L}\boldsymbol{B}}{M}
		\right]\\
		&= j2\pi\alpha(\boldsymbol{B}^H\boldsymbol{B})
		\left[
		-\frac{\bar{k}}{N},
		\frac{\bar{m}}{M}
		\right],
	\end{aligned}
	\label{eq:app_single_gain_coupling}
\end{equation}
where $\boldsymbol{B}^H\boldsymbol{B}=\sum_{k,m}p_{k,m}$.

Since $\boldsymbol{K}$ and $\boldsymbol{L}$ are diagonal with entries $k$ and $m$, respectively, and the squared entry magnitudes of $\boldsymbol{B}$ equal $p_{k,m}$, we have
\begin{equation*}
	\begin{aligned}
		\boldsymbol{B}^H\boldsymbol{K}^2\boldsymbol{B}
		&= \sum_{k,m}p_{k,m}k^2,
		&
		\boldsymbol{B}^H\boldsymbol{K}\boldsymbol{L}\boldsymbol{B}
		&= \sum_{k,m}p_{k,m}km,\\
		\boldsymbol{B}^H\boldsymbol{L}\boldsymbol{K}\boldsymbol{B}
		&= \sum_{k,m}p_{k,m}km,
		&
		\boldsymbol{B}^H\boldsymbol{L}^2\boldsymbol{B}
		&= \sum_{k,m}p_{k,m}m^2.
	\end{aligned}
\end{equation*}
The product $\boldsymbol{D}^H\boldsymbol{D}$ therefore admits the expression
\begin{equation}
	\begin{aligned}
		\boldsymbol{D}^H\boldsymbol{D}
		&= 4\pi^2|\alpha|^2
		\begin{bmatrix}
			\dfrac{\boldsymbol{B}^H\boldsymbol{K}^2\boldsymbol{B}}{N^2}
			&
			-\dfrac{\boldsymbol{B}^H\boldsymbol{K}\boldsymbol{L}\boldsymbol{B}}{NM}\\[4pt]
			-\dfrac{\boldsymbol{B}^H\boldsymbol{L}\boldsymbol{K}\boldsymbol{B}}{NM}
			&
			\dfrac{\boldsymbol{B}^H\boldsymbol{L}^2\boldsymbol{B}}{M^2}
		\end{bmatrix}\\
		&= 4\pi^2|\alpha|^2
		\sum_{k,m}p_{k,m}
		\begin{bmatrix}
			k^2/N^2 & -km/(NM)\\
			-km/(NM) & m^2/M^2
		\end{bmatrix}.
	\end{aligned}
	\label{eq:app_single_derivative_product}
\end{equation}
Substituting~\eqref{eq:gain_projector}, \eqref{eq:app_single_gain_coupling}, and~\eqref{eq:app_single_derivative_product} into~\eqref{eq:projected_fim} yields
\begin{equation}
	\begin{aligned}
		\boldsymbol{J}_{\mathrm e}(\boldsymbol{S})
		&= \frac{2}{\sigma_W^2}\operatorname{Re}
		\left\{
		\boldsymbol{D}^H\boldsymbol{D}
		-
		\frac{(\boldsymbol{B}^H\boldsymbol{D})^H(\boldsymbol{B}^H\boldsymbol{D})}
		{\boldsymbol{B}^H\boldsymbol{B}}
		\right\}\\
		&= \frac{8\pi^2|\alpha|^2}{\sigma_W^2}
		\begin{bmatrix}
			S_{kk}/N^2 & -S_{km}/(NM)\\
			-S_{km}/(NM) & S_{mm}/M^2
		\end{bmatrix}.
	\end{aligned}
	\label{eq:single_equivalent_fim}
\end{equation}
The subtracted coupling term converts the weighted second moments into the central moments defined in~\eqref{eq:centered_moments}. Thus, the effect of the unknown gain is accounted for through the energy-weighted centering of the subcarrier and slow-time indices.

For $\mathcal{D}=S_{kk}S_{mm}-S_{km}^2>0$, inversion gives
\begin{equation}
	\boldsymbol{J}_{\mathrm e}^{-1}(\boldsymbol{S})
	= \frac{\sigma_W^2}{8\pi^2|\alpha|^2\mathcal{D}}
	\begin{bmatrix}
		N^2S_{mm} & NM S_{km}\\
		NM S_{km} & M^2S_{kk}
	\end{bmatrix}.
	\label{eq:app_single_inverse}
\end{equation}
The diagonal entries are the normalized delay and Doppler CRBs with the unknown gain accounted for. Scaling by $\Delta R^2$ and $\Delta v^2$, respectively, gives the general-frame bounds in~\eqref{eq:single_range_velocity_crbs}.

For constant-modulus transmission, $p_{k,m}=E_s$, so the weighted centers are $\bar{k}=(N-1)/2$ and $\bar{m}=(M-1)/2$. The central moments reduce to
\begin{equation}
	\begin{aligned}
		S_{kk} &= \frac{E_sMN(N^2-1)}{12},\\
		S_{mm} &= \frac{E_sNM(M^2-1)}{12},\\
		S_{km} &= 0.
	\end{aligned}
	\label{eq:constant_moments}
\end{equation}
Substituting~\eqref{eq:constant_moments} into~\eqref{eq:single_range_velocity_crbs} and using $\Delta R=c/(2N\Delta f)$, $\Delta v=\lambda/(2MT_{\mathrm{sym}})$, and $\rho=|\alpha|^2E_s/\sigma_W^2$ gives the closed forms in~\eqref{eq:constant_range_velocity_crbs}. 

\section{Proof of Theorem~\ref{thm:qam_psk_crb_gap}}
\label{app:qam_psk_crb_gap}

Under PSK signaling, the conditional FIM equals $\overline{\boldsymbol{J}}$ for every frame. Under QAM signaling, fluctuations in symbol energy perturb the conditional FIM about the same mean. 
\subsection{CRB Increment Due to Symbol-Energy Fluctuations}
For $a\in\{\tau,\nu\}$, let $[\cdot]_{aa}$ denote the principal submatrix associated with the $Q$ delays or the $Q$ Doppler shifts. Taking this submatrix after inverting the full joint FIM accounts for the coupling among the unknown delays, Doppler shifts, and complex gains. By~\eqref{eq:average_crb} and~\eqref{eq:gap_relative_crb},
\begin{equation}
	\Gamma_a
	=
	\frac{
		\operatorname{Tr}\!\left(
		[\mathbb{E}[\boldsymbol{J}(\boldsymbol{S})^{-1}]-\overline{\boldsymbol{J}}^{-1}]_{aa}
		\right)
	}{
		\operatorname{Tr}\!\left([\overline{\boldsymbol{J}}^{-1}]_{aa}\right)
	},
	\label{eq:gap_joint_crb_ratio}
\end{equation}
where the expectation is over the QAM symbols.

To express the CRB increment relative to the PSK benchmark, define the normalized conditional FIM $\boldsymbol{T}=\overline{\boldsymbol{J}}^{-1/2}\boldsymbol{J}(\boldsymbol{S})\overline{\boldsymbol{J}}^{-1/2}$. Substituting $\boldsymbol{J}(\boldsymbol{S})^{-1}=\overline{\boldsymbol{J}}^{-1/2}\boldsymbol{T}^{-1}\overline{\boldsymbol{J}}^{-1/2}$ into~\eqref{eq:gap_joint_crb_ratio} and factoring the denominator in the same coordinates gives
\begin{equation}
	\Gamma_a
	=
	\frac{
		\operatorname{Tr}\!\left(
		[\overline{\boldsymbol{J}}^{-1/2}(\mathbb{E}[\boldsymbol{T}^{-1}]-\boldsymbol{I})\overline{\boldsymbol{J}}^{-1/2}]_{aa}
		\right)
	}{
		\operatorname{Tr}\!\left(
		[\overline{\boldsymbol{J}}^{-1/2}\boldsymbol{I}\,\overline{\boldsymbol{J}}^{-1/2}]_{aa}
		\right)
	}.
	\label{eq:gap_normalized_crb_ratio}
\end{equation}
It thus suffices to bound $\mathbb{E}[\boldsymbol{T}^{-1}]-\boldsymbol{I}$ relative to $\boldsymbol{I}$.

Let $K=NM$ be the number of resource elements in the OFDM frame. For resource element $(k,m)$, let $i=mN+k+1$ and $w_i=|S[k,m]|^2/E_s$. Let $\boldsymbol{g}_i\in\mathbb{C}^{1\times4Q}$ be the corresponding row of $\boldsymbol{G}_0$, and write $\boldsymbol{B}_i=\operatorname{Re}\{\boldsymbol{g}_i^H\boldsymbol{g}_i\}$. Expanding the FIM factorization in~\eqref{eq:energy_weighted_fim} over the resource elements gives
\begin{equation}
	\boldsymbol{J}(\boldsymbol{S})
	= \frac{2E_s}{\sigma_W^2}\sum_{i=1}^{K}w_i\boldsymbol{B}_i,
	\quad
	\boldsymbol{J}_0 = \frac{1}{K}\sum_{i=1}^{K}\boldsymbol{B}_i.
	\label{eq:gap_resource_fim}
\end{equation}
Thus, $\boldsymbol{B}_i$ gives the information contribution of resource element $i$ up to its energy and noise factor, and $\boldsymbol{J}_0$ averages these contributions over the frame. Substituting~\eqref{eq:gap_normalized_mean_fim} and~\eqref{eq:gap_resource_fim} into the definition of $\boldsymbol{T}$ yields
\begin{equation}
	\boldsymbol{T} = \sum_{i=1}^{K}w_i\boldsymbol{A}_i,
	\quad
	\boldsymbol{A}_i = \frac{1}{K}\boldsymbol{J}_0^{-1/2}\boldsymbol{B}_i\boldsymbol{J}_0^{-1/2}.
	\label{eq:gap_resource_weights}
\end{equation}
Since $\boldsymbol{A}_i\succeq\boldsymbol{0}$ and $\sum_i\boldsymbol{A}_i=\boldsymbol{I}$, the symbol-energy bounds imply that every frame satisfies
\begin{equation}
	\begin{aligned}
		w_{\min}\boldsymbol{I} &\preceq \boldsymbol{T}\preceq w_{\max}\boldsymbol{I},\\
		\frac{1}{w_{\max}}\boldsymbol{I} &\preceq \boldsymbol{T}^{-1}\preceq\frac{1}{w_{\min}}\boldsymbol{I}.
	\end{aligned}
	\label{eq:gap_inverse_spectral_bounds}
\end{equation}

By~\eqref{eq:gap_resource_weights}, the normalized FIM perturbation is
\begin{equation}
	\boldsymbol{E} = \boldsymbol{T}-\boldsymbol{I}
	= \sum_{i=1}^{K}(w_i-1)\boldsymbol{A}_i.
	\label{eq:gap_fim_perturbation}
\end{equation}
Since $\mathbb{E}[w_i]=\mathbb{E}[|S[k,m]|^2]/E_s=1$, we have $\mathbb{E}[\boldsymbol{E}]=\boldsymbol{0}$. Taking expectations in the matrix inverse identity
\begin{equation}
	\boldsymbol{T}^{-1}
	= \boldsymbol{I}-\boldsymbol{E}+\boldsymbol{E}\boldsymbol{T}^{-1}\boldsymbol{E}
	\label{eq:gap_inverse_identity}
\end{equation}
therefore gives
\begin{equation}
	\mathbb{E}[\boldsymbol{T}^{-1}]-\boldsymbol{I}
	= \mathbb{E}[\boldsymbol{E}\boldsymbol{T}^{-1}\boldsymbol{E}].
	\label{eq:gap_expected_inverse}
\end{equation}
For each frame, $\boldsymbol{E}$ is real and symmetric. Applying a congruence transformation with $\boldsymbol{E}$ to the bounds on $\boldsymbol{T}^{-1}$ in~\eqref{eq:gap_inverse_spectral_bounds} and then taking expectations gives
\begin{equation}
	\frac{\mathbb{E}[\boldsymbol{E}^2]}{w_{\max}}
	\preceq \mathbb{E}[\boldsymbol{T}^{-1}]-\boldsymbol{I}
	\preceq \frac{\mathbb{E}[\boldsymbol{E}^2]}{w_{\min}}.
	\label{eq:gap_second_moment_bounds}
\end{equation}
The second moment of the normalized FIM perturbation follows from~\eqref{eq:gap_fim_perturbation} as
\begin{equation}
	\begin{aligned}
		\mathbb{E}[\boldsymbol{E}^2]
		&= \sum_{i=1}^{K}\sum_{\ell=1}^{K}
		\mathbb{E}[(w_i-1)(w_\ell-1)]\,\boldsymbol{A}_i\boldsymbol{A}_\ell\\
		&\overset{(a)}{=} \sum_{i=1}^{K}\mathbb{E}[(w_i-1)^2]\boldsymbol{A}_i^2
		\overset{(b)}{=} (\kappa-1)\sum_{i=1}^{K}\boldsymbol{A}_i^2,
	\end{aligned}
	\label{eq:gap_perturbation_second_moment}
\end{equation}
where $(a)$ follows from independence across resource elements and $\mathbb{E}[w_i-1]=0$, and $(b)$ follows from $\mathbb{E}[(w_i-1)^2]=\mathbb{E}[|S[k,m]|^4]/E_s^2-1=\kappa-1$. Substituting~\eqref{eq:gap_perturbation_second_moment} into~\eqref{eq:gap_second_moment_bounds} yields
\begin{equation}
	\frac{\kappa-1}{w_{\max}}\sum_{i=1}^{K}\boldsymbol{A}_i^2
	\preceq \mathbb{E}[\boldsymbol{T}^{-1}]-\boldsymbol{I}
	\preceq \frac{\kappa-1}{w_{\min}}\sum_{i=1}^{K}\boldsymbol{A}_i^2.
	\label{eq:gap_inverse_increment_bounds}
\end{equation}

\subsection{Dependence on the Frame Dimensions and Target Count}

For a fixed symbol distribution, the scalar factors in~\eqref{eq:gap_inverse_increment_bounds} are constant. It remains to determine how $\sum_i\boldsymbol{A}_i^2$ scales with $N$, $M$, and $Q$. Since $\sum_i\boldsymbol{A}_i=\boldsymbol{I}$, we first bound $\boldsymbol{A}_i^2$ relative to $\boldsymbol{A}_i$ for each resource element.

Form the real matrix
\begin{equation}
	\boldsymbol{U}_i
	= \big[(\operatorname{Re}\boldsymbol{g}_i)^T,\,(\operatorname{Im}\boldsymbol{g}_i)^T\big]
	\in\mathbb{R}^{4Q\times2}.
	\label{eq:gap_real_jacobian}
\end{equation}
Expanding $\operatorname{Re}\{\boldsymbol{g}_i^H\boldsymbol{g}_i\}$ gives $\boldsymbol{B}_i=\boldsymbol{U}_i\boldsymbol{U}_i^T$. Substituting this factorization into the definition of $\boldsymbol{A}_i$ in~\eqref{eq:gap_resource_weights} and squaring the resulting expression gives
\begin{equation}
	\begin{aligned}
		\boldsymbol{A}_i
		&= \frac{1}{K}\boldsymbol{J}_0^{-1/2}\boldsymbol{U}_i\boldsymbol{I}_2
		\boldsymbol{U}_i^T\boldsymbol{J}_0^{-1/2},\\
		\boldsymbol{A}_i^2
		&= \frac{1}{K}\boldsymbol{J}_0^{-1/2}\boldsymbol{U}_i
		\left(\frac{1}{K}\boldsymbol{U}_i^T\boldsymbol{J}_0^{-1}\boldsymbol{U}_i\right)
		\boldsymbol{U}_i^T\boldsymbol{J}_0^{-1/2}.
	\end{aligned}
	\label{eq:gap_resource_square}
\end{equation}
The common outer factors reduce the comparison to bounding $K^{-1}\boldsymbol{U}_i^T\boldsymbol{J}_0^{-1}\boldsymbol{U}_i$ relative to $\boldsymbol{I}_2$. The gain derivatives provide the lower bound, while all four parameter groups enter the upper bound.

At resource element $(k,m)$, write the channel response as $h_i=\boldsymbol{a}_i^T\boldsymbol{\alpha}$, where $\boldsymbol{a}_i=[a_{i,1},\ldots,a_{i,Q}]^T$ and
\begin{equation}
	a_{i,q}=e^{-j2\pi k\tau_q/N}e^{j2\pi m\nu_q/M}.
	\label{eq:gap_steering_coefficient}
\end{equation}
The four parameter groups in the Jacobian row of~\eqref{eq:channel_jacobian_meaning} have derivatives
\begin{equation}
	\begin{aligned}
		\frac{\partial h_i}{\partial\boldsymbol{\tau}^T}
		&= -\frac{j2\pi k}{N}(\boldsymbol{\alpha}\odot\boldsymbol{a}_i)^T,\\
		\frac{\partial h_i}{\partial\boldsymbol{\nu}^T}
		&= \frac{j2\pi m}{M}(\boldsymbol{\alpha}\odot\boldsymbol{a}_i)^T,\\
		\frac{\partial h_i}{\partial\boldsymbol{\alpha}_R^T}
		&= \boldsymbol{a}_i^T,
		\quad
		\frac{\partial h_i}{\partial\boldsymbol{\alpha}_I^T}
		= j\boldsymbol{a}_i^T.
	\end{aligned}
	\label{eq:gap_channel_derivatives}
\end{equation}
Separating the real and imaginary parts as in~\eqref{eq:gap_real_jacobian} gives
\begin{equation}
	\boldsymbol{U}_i
	=
	\begin{bmatrix}
		\dfrac{2\pi k}{N}\operatorname{Im}(\boldsymbol{\alpha}\odot\boldsymbol{a}_i)
		&
		-\dfrac{2\pi k}{N}\operatorname{Re}(\boldsymbol{\alpha}\odot\boldsymbol{a}_i)\\[2pt]
		-\dfrac{2\pi m}{M}\operatorname{Im}(\boldsymbol{\alpha}\odot\boldsymbol{a}_i)
		&
		\dfrac{2\pi m}{M}\operatorname{Re}(\boldsymbol{\alpha}\odot\boldsymbol{a}_i)\\[2pt]
		\operatorname{Re}\boldsymbol{a}_i & \operatorname{Im}\boldsymbol{a}_i\\
		-\operatorname{Im}\boldsymbol{a}_i & \operatorname{Re}\boldsymbol{a}_i
	\end{bmatrix}.
	\label{eq:gap_real_jacobian_blocks}
\end{equation}

Each row block in~\eqref{eq:gap_real_jacobian_blocks} contributes a positive semidefinite term to $\boldsymbol{U}_i^T\boldsymbol{U}_i$. Retaining the two gain blocks gives
\begin{equation}
	\begin{aligned}
		\boldsymbol{U}_i^T\boldsymbol{U}_i
		&\succeq
		\begin{bmatrix}
			\operatorname{Re}\boldsymbol{a}_i & \operatorname{Im}\boldsymbol{a}_i\\
			-\operatorname{Im}\boldsymbol{a}_i & \operatorname{Re}\boldsymbol{a}_i
		\end{bmatrix}^{T}
		\begin{bmatrix}
			\operatorname{Re}\boldsymbol{a}_i & \operatorname{Im}\boldsymbol{a}_i\\
			-\operatorname{Im}\boldsymbol{a}_i & \operatorname{Re}\boldsymbol{a}_i
		\end{bmatrix}\\
		&= \|\boldsymbol{a}_i\|_2^2\boldsymbol{I}_2 = Q\boldsymbol{I}_2.
	\end{aligned}
	\label{eq:gap_gain_gram_lower}
\end{equation}
The factor $Q$ follows from the unit modulus of each steering coefficient. For target $q$, the gain derivatives $a_{i,q}$ and $ja_{i,q}$ together contribute $|a_{i,q}|^2\boldsymbol{I}_2=\boldsymbol{I}_2$.

For the upper bound, summing the squared Frobenius norms of all four blocks in~\eqref{eq:gap_real_jacobian_blocks} gives
\begin{equation}
	\begin{aligned}
		\lambda_{\max}(\boldsymbol{U}_i^T\boldsymbol{U}_i)
		&\leq \operatorname{Tr}(\boldsymbol{U}_i^T\boldsymbol{U}_i)\\
		&= 2Q+4\pi^2\left(\frac{k^2}{N^2}+\frac{m^2}{M^2}\right)\|\boldsymbol{\alpha}\|_2^2\\
		&\leq (2+8\pi^2\alpha_{\max}^2)Q,
	\end{aligned}
	\label{eq:gap_gram_trace_upper}
\end{equation}
where $0\leq k<N$, $0\leq m<M$, and $\|\boldsymbol{\alpha}\|_2^2\leq Q\alpha_{\max}^2$. Since
\begin{equation*}
	\frac{\boldsymbol{U}_i^T\boldsymbol{U}_i}{\lambda_{\max}(\boldsymbol{J}_0)}
	\preceq \boldsymbol{U}_i^T\boldsymbol{J}_0^{-1}\boldsymbol{U}_i
	\preceq \frac{\boldsymbol{U}_i^T\boldsymbol{U}_i}{\lambda_{\min}(\boldsymbol{J}_0)},
\end{equation*}
combining~\eqref{eq:gap_gain_gram_lower} and~\eqref{eq:gap_gram_trace_upper} yields
\begin{equation}
	\frac{Q}{K\lambda_{\max}(\boldsymbol{J}_0)}\boldsymbol{I}_2
	\preceq \frac{1}{K}\boldsymbol{U}_i^T\boldsymbol{J}_0^{-1}\boldsymbol{U}_i
	\preceq \frac{(2+8\pi^2\alpha_{\max}^2)Q}{K\lambda_{\min}(\boldsymbol{J}_0)}\boldsymbol{I}_2.
	\label{eq:gap_weighted_gram_bounds}
\end{equation}

Applying~\eqref{eq:gap_weighted_gram_bounds} to the factorization of $\boldsymbol{A}_i^2$ in~\eqref{eq:gap_resource_square} and using the corresponding factorization of $\boldsymbol{A}_i$ gives
\begin{equation}
	\frac{Q\,\boldsymbol{A}_i}{K\lambda_{\max}(\boldsymbol{J}_0)}
	\preceq \boldsymbol{A}_i^2
	\preceq \frac{(2+8\pi^2\alpha_{\max}^2)Q\,\boldsymbol{A}_i}{K\lambda_{\min}(\boldsymbol{J}_0)}.
	\label{eq:gap_resource_comparison}
\end{equation}
Summing over $i$ and using $\sum_i\boldsymbol{A}_i=\boldsymbol{I}$, we obtain
\begin{equation}
	\frac{Q}{K\lambda_{\max}(\boldsymbol{J}_0)}\boldsymbol{I}
	\preceq \sum_{i=1}^{K}\boldsymbol{A}_i^2
	\preceq \frac{(2+8\pi^2\alpha_{\max}^2)Q}{K\lambda_{\min}(\boldsymbol{J}_0)}\boldsymbol{I}.
	\label{eq:gap_sum_resource_squares}
\end{equation}

Finally, substituting~\eqref{eq:gap_sum_resource_squares} into~\eqref{eq:gap_inverse_increment_bounds} and applying a congruence transformation with $\overline{\boldsymbol{J}}^{-1/2}$ yields the bounds on the full joint CRB increment,
\begin{equation}
	\begin{aligned}
		\frac{(\kappa-1)Q\,\overline{\boldsymbol{J}}^{-1}}{w_{\max}K\lambda_{\max}(\boldsymbol{J}_0)}
		&\preceq \mathbb{E}[\boldsymbol{J}(\boldsymbol{S})^{-1}]-\overline{\boldsymbol{J}}^{-1}\\
		&\preceq \frac{(\kappa-1)(2+8\pi^2\alpha_{\max}^2)Q\,\overline{\boldsymbol{J}}^{-1}}{w_{\min}K\lambda_{\min}(\boldsymbol{J}_0)}.
	\end{aligned}
	\label{eq:gap_full_joint_crb_bounds}
\end{equation}
Taking the trace of the delay or Doppler principal submatrix in~\eqref{eq:gap_full_joint_crb_bounds} and dividing by the corresponding PSK total CRB, $\operatorname{Tr}([\overline{\boldsymbol{J}}^{-1}]_{aa})$, gives~\eqref{eq:gap_finite_frame_bounds} by~\eqref{eq:gap_joint_crb_ratio} and $K=NM$.

Under the stated uniform bounds, the coefficient of $Q/(NM)$ in the lower bound of~\eqref{eq:gap_finite_frame_bounds} is bounded away from zero, and the corresponding coefficient in the upper bound is bounded above. This proves~\eqref{eq:gap_asymptotic_order}. The range and velocity conversion factors cancel in their respective ratios, completing the proof.

\bibliographystyle{ieeetran}
\bibliography{ref_abbreviation}	

\begin{IEEEbiography}[{\includegraphics[width=1in,height=1.25in,clip,keepaspectratio]{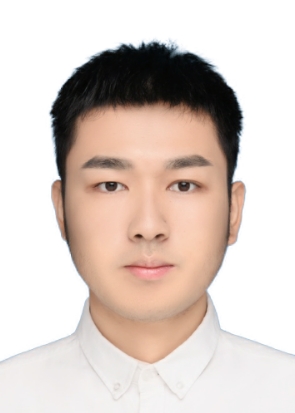}}]{Junjie~Chen} received the B.Eng. degree from the Beijing Institute of Technology (BIT), Beijing, China, in 2017, and the M.E. degree from the University of Science and Technology of China (USTC), Hefei, China, in 2020. 
From 2020 to 2022, he was with the China Electronics Technology Group Corporation (CETC), where he conducted research on radar and communication systems. 
He received the Ph.D. degree from Sun Yat-sen University (SYSU), Guangzhou, China, in 2026. 
He is currently a lecturer with the School of Computing and Artificial Intelligence, Jiangxi University of Finance and Economics, Nanchang, China. 
His research interests include signal processing, wireless communications, and machine learning. \end{IEEEbiography}

\end{document}